\documentclass[lettersize,journal]{IEEEtran}

\usepackage{xspace}
\usepackage{subfigure}
\usepackage{graphicx}
\usepackage{balance}
\usepackage{comment}
\usepackage{alltt}
\usepackage{multirow}
\usepackage{CJK}
\usepackage{textcomp}
\usepackage{colortbl}
\usepackage{amsmath}
\usepackage{amsfonts}
\usepackage{url}
\usepackage{algorithmic}
\usepackage[ruled,vlined]{algorithm2e}
\usepackage{multicol}
\usepackage{lipsum}
\usepackage{mathrsfs}
\usepackage{overpic}
\usepackage{booktabs}
\usepackage{bbm}
\usepackage{makecell}
\usepackage{amsthm}

\newcommand{\ie}{\emph{i.e.}\xspace}
\newcommand{\eg}{\emph{e.g.}\xspace}
\newcommand{\etc}{\emph{etc}\xspace}

\newcommand{\Figure}[1]{Fig.~\ref{fig:#1}}
\newcommand{\Equation}[1]{Eq.~\ref{eq:#1}}
\newcommand{\Section}[1]{\S\ref{sec:#1}}
\newcommand{\Table}[1]{Table~\ref{tab:#1}}

\newcommand\system{mmADA\xspace}

\begin{document}

% \title{Active Domain Adaptation for mmWave-based Human Activity Recognition via Sample Uncertainty Estimation}
\title{Active Domain Adaptation for mmWave-based HAR via R$\acute{e}$nyi Entropy-based Uncertainty Estimation}

% \author{Anonymous Authors}
\author{Mingzhi Lin, Teng Huang, Han Ding*, Cui Zhao, Fei Wang, Ge Wang, Wei Xi%
\thanks{Our manuscript is currently under review.} %
\thanks{Mingzhi Lin, Teng Huang, Han Ding, Cui Zhao, Ge Wang, and Wei Xi are with the School of Computer Science and Engineering, Xi'an Jiaotong University, China.} %
\thanks{Fei Wang is with the School of Software Engineering, Xi'an Jiaotong University, China.} %
\thanks{* Han Ding is the Corresponding author.} %
\thanks{Digital Object Identifier xxx/xxx.}}

\markboth{Journal of \LaTeX\ Class Files,~Vol.~14, No.~8, August~2025}%
{Shell \MakeLowercase{\textit{et al.}}: A Sample Article Using IEEEtran.cls for IEEE Journals}

\maketitle

\begin{abstract}
Human Activity Recognition (HAR) using mmWave radar provides a non-invasive alternative to traditional sensor-based methods but suffers from domain shift, where model performance declines in new users, positions, or environments. To address this, we propose \system, an Active Domain Adaptation (ADA) framework that efficiently adapts mmWave-based HAR models with minimal labeled data. \system enhances adaptation by introducing R$\acute{e}$nyi Entropy-based uncertainty estimation to identify and label the most informative target samples. Additionally, it leverages contrastive learning and pseudo-labeling to refine feature alignment using unlabeled data. Evaluations with a TI IWR1443BOOST radar across multiple users, positions, and environments show that \system achieves over 90\% accuracy in various cross-domain settings. Comparisons with five baselines confirm its superior adaptation performance, while further tests on unseen users, environments, and two additional open-source datasets validate its robustness and generalization.  
\end{abstract}


\section{Introduction}

Human Activity Recognition (HAR) is a fundamental technology with applications in health monitoring, smart homes, human-computer interaction, \etc. Traditional HAR methods primarily rely on wearable sensors or cameras, which can be intrusive, uncomfortable, or raise privacy concerns. Wireless HAR, which leverages radio frequency signals to detect and classify human activities, offers a non-invasive alternative. Technologies such as Wi-Fi \cite{yan2024person}\cite{zhao2024one}\cite{zhou2022target}, RFID \cite{ding2015femo}\cite{zhao2023rfid}, and ultrasound \cite{cao2023can}\cite{wang2024amt}, have been explored for wireless HAR by capturing motion through signal reflections or time-of-flight measurements. While promising, these methods often struggle with accuracy in complex environments and are highly susceptible to interference.
Millimeter-wave (mmWave) radar, in contrast, provides high-resolution spatial and temporal information, enabling precise motion detection even in challenging conditions \cite{lu2020see}\cite{zhu2023tilemask}. Despite its advantages, deploying mmWave-based HAR models in real-world scenarios is still hindered by \textit{domain shift}—a significant drop in model performance when applied to agnostic new domains, including new environments, positions, or users, that differ from those seen during training.

To mitigate domain shift, researchers have explored various strategies. 
Early efforts focused on designing signal processing features that are more domain-invariant, such as Time-of-Flight (ToF), Doppler Frequency Shift (DFS), and Body-coordinate Velocity Profile (BVP) \cite{Widar2.0}\cite{widar3.0}.
More recent approaches incorporate domain adaptation techniques, including adversarial learning, meta-learning, and few-shot learning. 
Adversarial learning \cite{jiang2018towards}\cite{zhang2021privacy} mitigates domain-specific biases by training a discriminator to align features across domains, but its effectiveness is limited by the diversity of available source domains and unstable training.
Meta-learning \cite{ding2020rf}\cite{zhang2024few} aims to learn transferable knowledge across tasks, enabling rapid adaptation, but it often fails when the source and target domains differ significantly and can be computationally expensive.
Few-shot learning \cite{feng2022wi}\cite{gong2019metasense}\cite{xia2024ts2act}\cite{xiao2021onefi} fine-tunes a pre-trained model using only a few labeled samples from the target domain, reducing label requirements. Yet, it can overfit to these few samples. In addition, it typically distributes labels equally across all activity classes, regardless of their actual impact on performance.
This raises a critical question: \textit{Is equal label allocation across classes the most efficient way to spend a limited labeling budget?} We argue that selectively labeling only the most valuable target samples is a more effective strategy, especially in cross-domain HAR, where certain activities may require more targeted supervision than others.

To this end, we explore active learning as a principled solution. Active learning prioritizes the annotation of samples that are most uncertain or informative, thereby improving model performance while minimizing labeling effort.
Building on this, we propose \system, an Active Domain Adaptation (ADA) framework that efficiently adapts mmWave-based HAR models to various domains with minimal labeled data. Our approach addresses two key challenges:
(i) \textbf{\textit{How to select the most informative target samples under agnostic domain variations?}} Since labeling in wireless sensing is costly and time-consuming, \system aims to minimize annotation requirements while ensuring effective adaptation. Rather than selecting samples randomly, we propose an uncertainty-driven sampling strategy based on Evidential Deep Learning (EDL). Unlike prior methods \cite{prabhu2021active}\cite{xie2023dirichlet}\cite{zhang2024revisiting} that rely on Shannon Entropy for uncertainty estimation, we are the first to introduce R$\acute{e}$nyi Entropy in this context. This allows \system to focus more precisely on hard-to-classify samples, particularly those affected by severe signal attenuation. We derive formulations for domain uncertainty (measuring domain shift) and prediction uncertainty (quantifying model confidence) under a Dirichlet prior using R$\acute{e}$nyi Entropy. By jointly capturing these two uncertainties, \system reliably identifies the most valuable samples for labeling, driving more effective domain adaptation.
(ii) \textbf{\textit{How to exploit the knowledge of the remaining unlabeled target data for enhanced adaptation?}} We observe that samples from the same activity class naturally cluster in the latent space, while samples from different classes remain well separated. Leveraging this property, we integrate contrastive learning to improve class-level alignment between the source and target domains. Additionally, we introduce an adaptive pseudo-labeling strategy that utilizes a KNN-based similarity metric to assign pseudo labels to unlabeled target samples, further refining the adaptation process and boosting model performance.

We evaluate \system with a commodity off-the-shelf TI IWR1443BOOST radar. The experiments involve 8 users, 5 positions, 2 environments, and 11 predefined activities. Our approach achieves over 90\% recognition accuracy under cross-user, cross-position, and cross-environment conditions, \textit{using only two users' data from one position as the source domain}. Comparisons with five representative baseline methods \cite{liu2022mtranssee}\cite{jiang2018towards}\cite{zhang2024few}\cite{xie2022active}\cite{zhang2024revisiting} demonstrate \system's superiority. Furthermore, evaluations on unseen users, environments, and two additional open-source datasets validate its robustness and generalization.

We summarize the contributions as follows:

(i) We introduce \system, the first active domain adaptation framework designed for mmWave-based HAR, which effectively mitigates performance degradation when adapting to new domains with minimal labeling costs.

(ii) We are the first to propose a novel R$\acute{e}$nyi Entropy-based uncertainty estimation method that simultaneously captures domain uncertainty and prediction uncertainty, enabling \system to selectively identify and label the most informative target samples for improved learning.

(iii) We conduct extensive experiments across diverse cross-domain settings, where \system consistently outperforms five baseline methods. Besides, \system also achieves superior performance on two newly introduced large-scale open-source datasets, XRF55 \cite{wang2024xrf55} and MM-Fi \cite{yang2023mm}.

\section{mmWave Sensing Principles}\label{sec:process}

% \subsection{Sensing Principles}

% The mmWave radar operates within the 30 GHz to 300 GHz frequency range, offering high resolution for sensing applications. A commonly used architecture is the Frequency Modulated Continuous Wave (FMCW) radar, which transmits a series of chirp signals with linearly swept frequencies. Each chirp is characterized by a start frequency $f_0$, sweep bandwidth $B$, and chirp duration $T_c$. The radar forms a frame by combining multiple chirps, and frames are transmitted at a high rate, enabling real-time sensing. The round-trip delay between the transmitted and received signals creates a frequency shift in the intermediate frequency (IF) signal, which serves as the basis for extracting range, velocity, and angular information about objects within the field of view.

% \subsection{mmWave Sensing Signatures}\label{sec:process}
The FMCW mmWave radar transmits a series of chirp signals with linearly swept frequencies. The round-trip delay between the transmitted and received signals creates a frequency shift in the intermediate frequency (IF) signal.
The IF signal, a complex time-domain waveform, undergoes a series of signal processing steps to extract meaningful signatures for wireless sensing applications.

\textit{Range-Doppler and Range-Angle maps}. To isolate objects at different distances, a range-FFT is performed along the fast-time axis of each chirp. Subsequently, a Doppler-FFT along the slow-time axis (across chirps) reveals the relative velocity of objects, producing the Range-Doppler map. 
%Modern mmWave radar, equipped with multiple transmitting and receiving antennas, allows for spatial processing. 
%
Besides, an angle-FFT on the signals received across the antenna elements generates the Range-Angle map, providing insights into the azimuthal positions of detected objects.

\textit{Time-Doppler heatmap}. 
Human activities inherently exhibit temporal continuity, spanning multiple time frames. 
To effectively capture long-range temporal dependencies, we extract sequential features.
Specifically, we generate the Time-Doppler (TD) heatmap by summing the Range-Doppler heatmap along the range dimension and integrating the resulting vectors across consecutive frames. This representation enhances activity characterization by preserving temporal motion patterns.
%
% The TD heatmap reveals the pattern of human activity. Focusing on the dynamic regions in the sequential heatmaps not only enhances interpretability but also improves feature discriminability for downstream neural networks.

\textit{Time-Angle heatmap}. 
Similar to Range-Doppler processing, analyzing Range-Angle maps across frames helps characterize spatial-temporal activity patterns. However, multipath reflections and static clutter introduce noise, requiring a filtering step to isolate human-related signals. 
To address this, we perform mean pooling across antenna elements for all Range-Doppler maps obtained from different antenna pairs. We then accumulate intensity values under various Doppler bins, retaining only those exceeding a predefined statistical threshold and mapping them to their corresponding Range-Angle indices. After noise elimination \cite{li2022towards}, the refined Range-Angle maps highlight human-centric regions, facilitating the construction of Time-Angle (TA) heatmaps.

\begin{figure}[t]
    \centering
    \includegraphics[width=.46\textwidth]{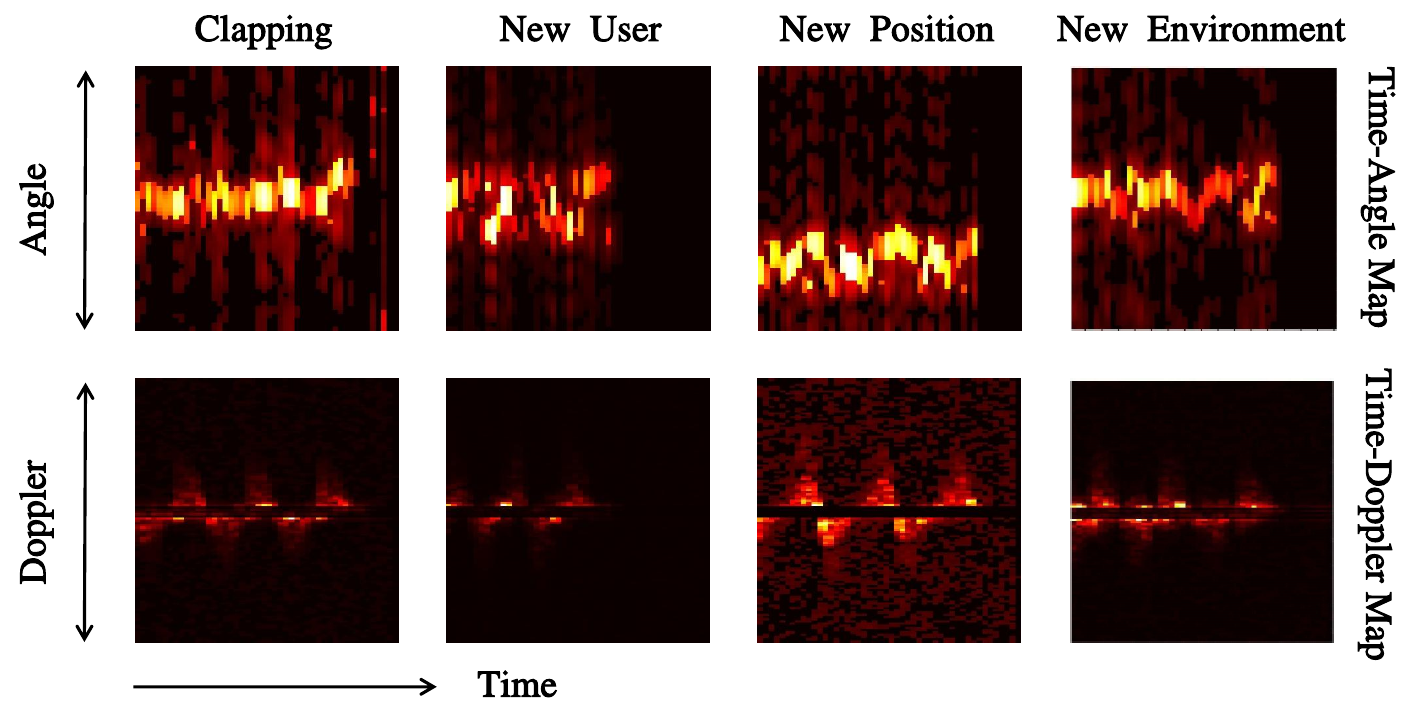}
    \caption{Time-Angle heatmaps and Time-Doppler heatmaps of a activity(clap) of four domains.}
    \label{fig:pre_TD_TA}
    \vspace{-0.1in}
\end{figure}

In our implementation, TD and TA heatmaps are jointly used to profile activities. The first column of \Figure{pre_TD_TA} shows the TD and TA of activity `clapping'. We can see that TD heatmap captures velocity variations, while TA heatmap depicts spatial movement patterns. The dynamic regions in these heatmaps enhance feature discriminability, benefiting downstream neural network processing.

\section{Motivation and Overview}

% \subsection{Problem Definition}

\subsection{Domain Shifts in mmWave-based HAR}
An mmWave-based HAR model trained in a source domain (\eg, a group of users in a fixed environment) often struggles to generalize to new domains due to three key factors:

i) \textit{New Positions:} The relative position between the human subject and the radar significantly impacts the received signal features. In the source domain, the model may learn position-dependent features, such as specific range and angle reflections. When deployed to new positions, these signatures change due to variations in incident angles and multipath effects, leading to degraded recognition performance. 

ii) \textit{New Users:} Human movements exhibit inter-person variability due to differences in height, limb lengths, and motion patterns. A model trained on a limited set of users may overfit to their unique characteristics.
%, rather than learning generalized representations of the activities. 
%
When the model encounters new users, it may misclassify their activities due to unfamiliar motion signatures.

\begin{figure}[t]
    \centering
    \includegraphics[width=.46\textwidth]{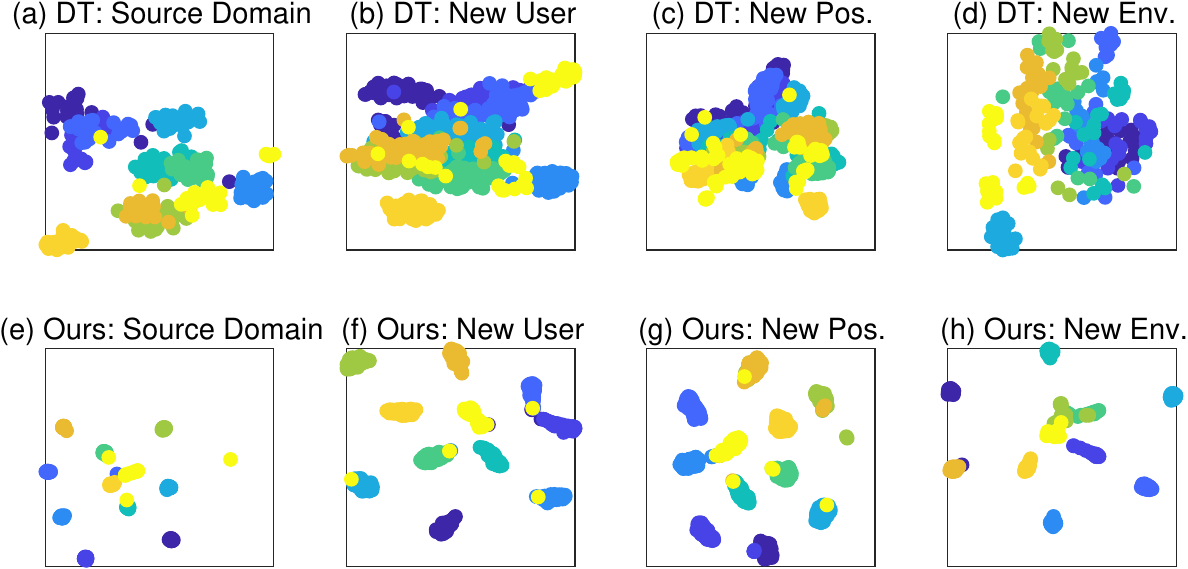}
    \caption{t-SNE visualization of feature embeddings.} 
    %(a)$\sim$(d) Direct Transfer (DT). (e)$\sim$(h) \system.
    \label{fig:pre_tsne}
    \vspace{-0.1in}
\end{figure}

% \begin{figure}[t]
%     \centering
%     \includegraphics[width=.46\textwidth]{pic/pre_tsne1}
%     \caption{t-SNE visualization of feature embeddings.} 
%     %(a)$\sim$(d) Direct Transfer (DT). (e)$\sim$(h) \system.
%     \label{fig:pre_tsne}
%     \vspace{-0.1in}
% \end{figure}

% \begin{figure*}[t]
% \centering
% \begin{minipage}[t]{\linewidth}
% \centering
% \subfigure[DT: Shannon Entropy]
% {\includegraphics[width=.242\textwidth]{pic/pre_DT_shannon}\label{DT shannon}}
% % \hspace{0.15in}
% \subfigure[DT: R$\acute{e}$nyi Entropy]
% {\includegraphics[width=.242\textwidth]{pic/pre_DT_renyi}\label{DT renyi}}
% \subfigure[Ours: Shannon Entropy]
% {\includegraphics[width=.242\textwidth]{pic/pre_ours_shannon}\label{Ours shannon}}
% % \hspace{0.15in}
% \subfigure[Ours: R$\acute{e}$nyi Entropy]
% {\includegraphics[width=.242\textwidth]{pic/pre_ours_renyi}\label{Ours renyi}}
% \caption{Kernel Density Estimation (KDE) of domain uncertainty across the source, target and unseen domains.}%(a)(b) Domain uncertainty measured by Shannon entropy and R$\acute{e}$nyi entropy under DT. (c)(d) Domain uncertainty of \system.
% \label{fig:pre_distribution}
% \end{minipage}
% % \vspace{-0.1in}
% \end{figure*}

iii) \textit{New Environments:} Environmental factors such as furniture layouts, and ambient interference cause substantial shifts in mmWave signal propagation. Reflections and multipath components in the source environment will change in the target environment, altering range, Doppler, and angle signatures. These environmental shifts often introduce noise that hinder model generalization.

\textbf{Proof of Domain Shifts.} As illustrated in \Figure{pre_TD_TA}, TA and TD heatmaps exhibit noticeable changes when exposed to new position, user, or environment, indicating the domain shift. To further demonstrate its impact on mmWave-based HAR, we conduct a series of proof-of-concept experiments. Specifically, we train a deep neural network—comprising a feature extractor and a classifier—on a source domain where users perform 11 activities. We then visualize the latent features extracted by the model using t-SNE, as shown in \Figure{pre_tsne}. 
In the source domain (\Figure{pre_tsne}(a)), activities form well-clustered groups with relatively clear boundaries between categories. However, when the same model is applied directly to new domains without adaptation (referred to as \textit{direct transfer} in the following), the features become scattered (\Figure{pre_tsne}(b)$\sim$(d)), highlighting the existence of domain shifts.

\begin{figure}[t]
\centering
\begin{minipage}[t]{\linewidth}
\centering
\subfigure[DT: Shannon Entropy]
{\includegraphics[width=.39\textwidth]{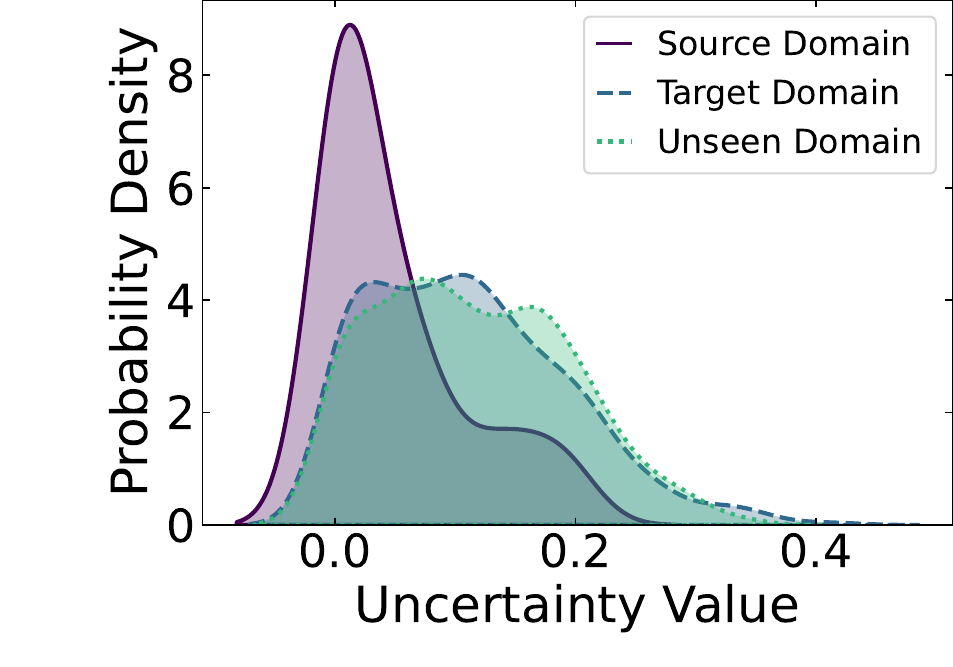}}
\hspace{0.15in}
\subfigure[DT: R$\acute{e}$nyi Entropy]
{\includegraphics[width=.38\textwidth]{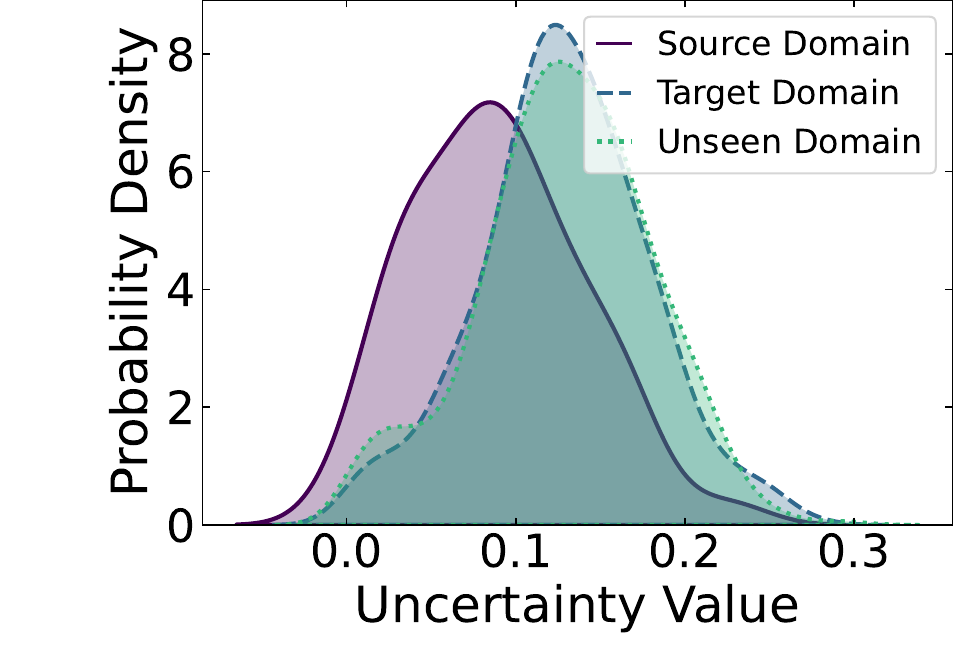}}
\subfigure[Ours: Shannon Entropy]
{\includegraphics[width=.38\textwidth]{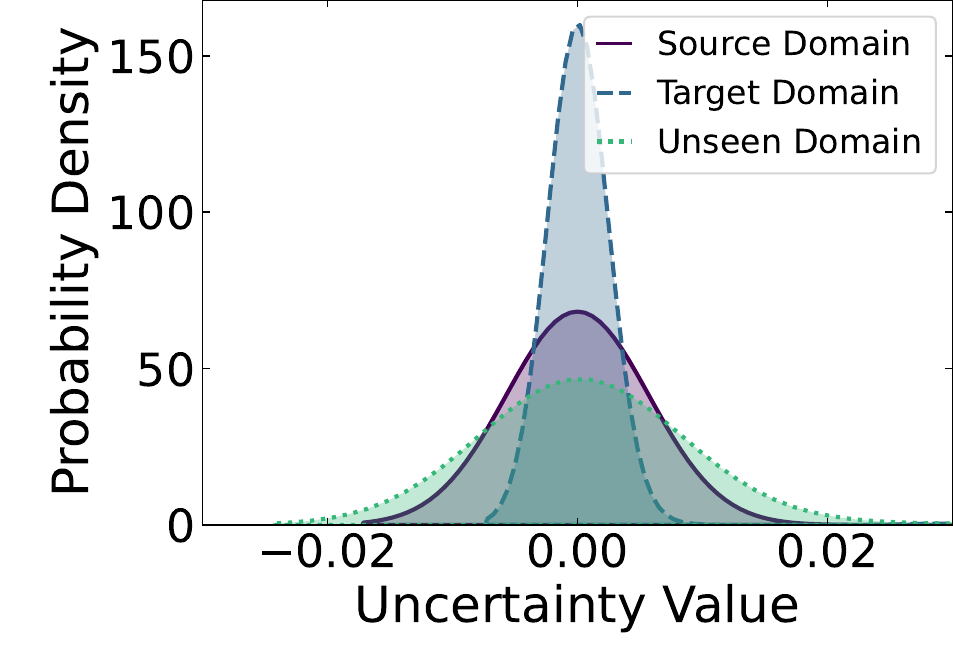}}
\hspace{0.15in}
\subfigure[Ours: R$\acute{e}$nyi Entropy]
{\includegraphics[width=.38\textwidth]{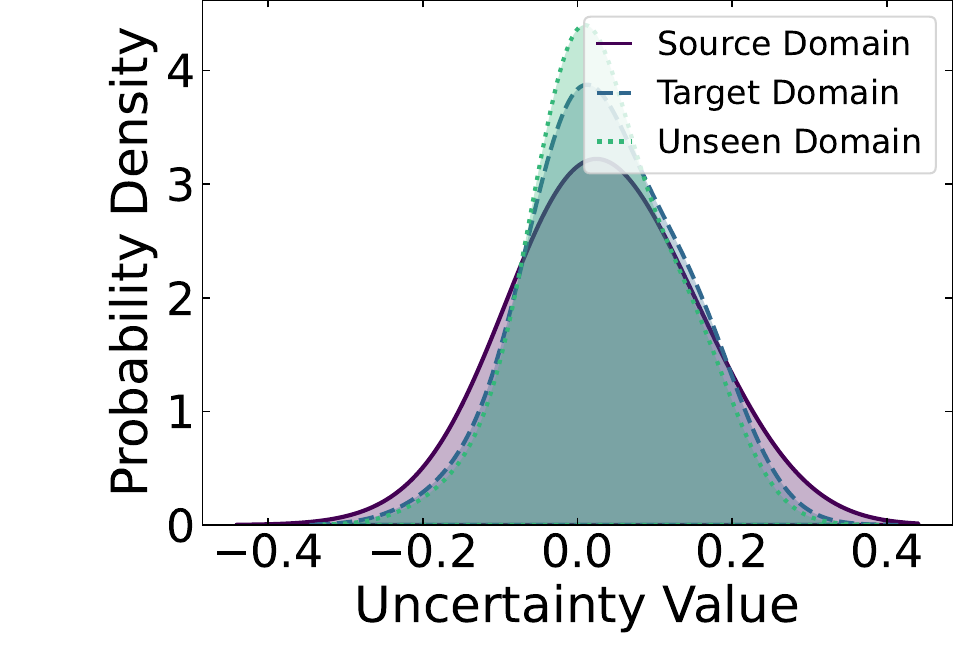}}
\caption{Kernel Density Estimation (KDE) of domain uncertainty across the source, target and unseen domains.}%(a)(b) Domain uncertainty measured by Shannon entropy and R$\acute{e}$nyi entropy under DT. (c)(d) Domain uncertainty of \system.
\label{fig:pre_distribution}
\end{minipage}
\vspace{-0.25in}
\end{figure}

To quantify this effect, we draw on prior work \cite{malinin2018predictive}, which introduces domain uncertainty to measure the mismatch between estimated and actual domain characteristics. \Figure{pre_distribution}(a) and (b) compare uncertainty distributions across the source, target, and unseen domains under direct transfer (DT). 
The source domain shows consistently low uncertainty, while both the target and unseen domains exhibit substantially higher uncertainty—further validating the domain shift.
Fortunately, our proposed \system framework can effectively mitigate domain shifts. As shown in \Figure{pre_distribution}(c) and (d), applying \system results in near-zero domain uncertainty across all domains, demonstrating its effectiveness in adapting to new domains.

\textbf{Shannon Entropy vs R$\acute{e}$nyi Entropy.}
From \Figure{pre_distribution}(a)(b) and (c)(d), we have two key observations. First, the uncertainty distributions of the target and unseen domains are more consistent when measured with R$\acute{e}$nyi entropy compared to Shannon entropy—under both direct transfer and our proposed method. This suggests that R$\acute{e}$nyi entropy may provide better generalization across domains. Second, after applying domain adaptation with R$\acute{e}$nyi entropy (\Figure{pre_distribution}(d)), the uncertainty distributions of both the target and unseen domains align closely with that of the source domain. This indicates that the adapted model perceives these domains similarly, effectively narrowing the domain gap. These findings highlight the value of R$\acute{e}$nyi entropy as a more informative uncertainty measure, capable of improving both adaptation performance and generalization. Additional empirical validation is presented in \Section{proof_renyi_over_shannon}.

\subsection{Overview}

\begin{figure*}[t]
    \centering
    \includegraphics[width=.85\textwidth]{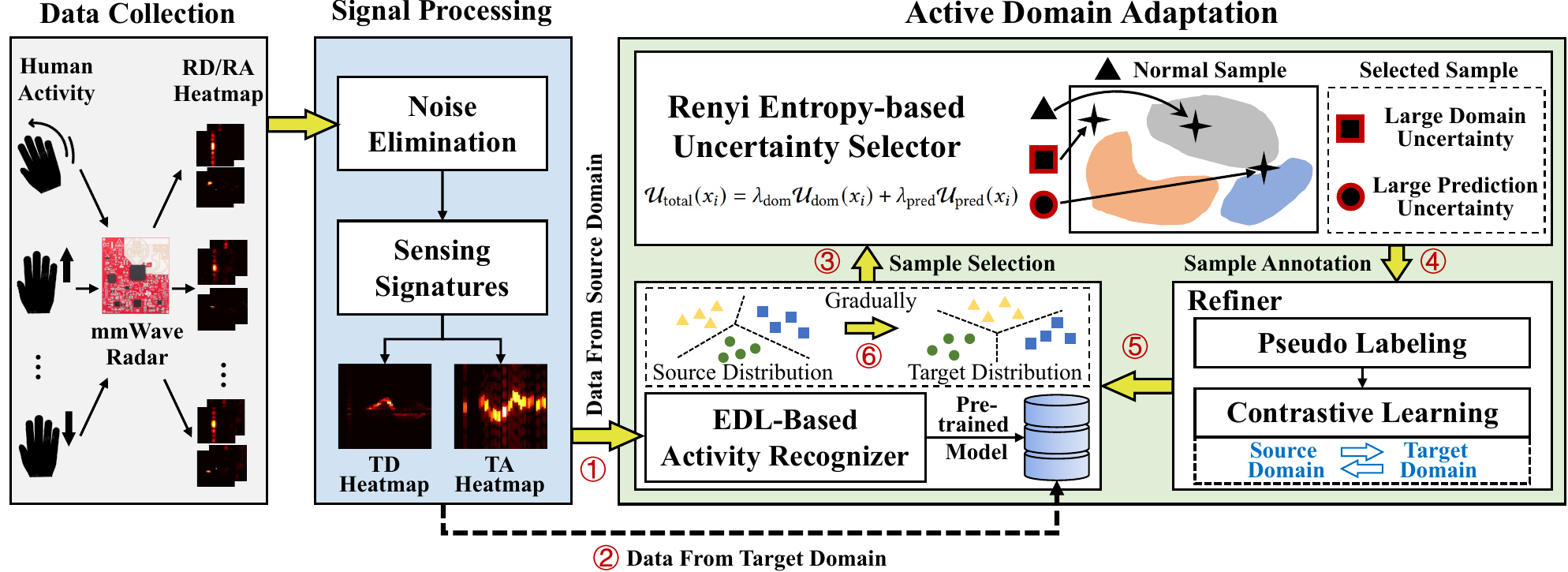}
    \caption{Overview of \system.}
    \label{fig:overview}
    \vspace{-0.10in}
\end{figure*}

In this paper, we introduce a method that facilitates the adaptation of the mmWave-based HAR system to an agnostic new domain with minimal effort. This is achieved by transferring the HAR model from the source to the target domain using active learning, which selects and labels the most informative samples from the new domain based on uncertainty estimation. As shown in \Figure{overview}, the workflow of our method is outlined below.

$\bullet$ \textbf{Signal Processing}. To represent different human activities, we first preprocess the mmWave signals to generate the corresponding TD and TA heatmaps.
Additionally, to reduce ambient reflections, we perform noise elimination as detailed in \Section{process}.

$\bullet$ \textbf{Recognition in the Source Domain}. We begin by proposing an activity recognition model based on Evidential Deep Learning (EDL) for the source domain. This model consists of a feature extractor and a classifier. The classifier uses Dirichlet priors on class probabilities, allowing us to interpret the classification predictions and estimate uncertainties.

$\bullet$ \textbf{Recognition in the Target Domain}. In this step, we adapt the HAR model to the target domain using active learning. Specifically, we employ R$\acute{e}$nyi Entropy to quantify both \textit{domain and prediction uncertainties}, selecting samples with the highest joint uncertainty for labeling. These newly labeled samples are then used to further train the model. Additionally, to maximize the value of unlabeled data, we integrate pseudo labeling and contrastive learning, enhancing feature alignment and overall model performance.

\section{System Design}

\subsection{Problem Definition}
Formally, we consider a fully labeled source domain and an unlabeled target domain, both containing the same set of activity classes. Let the source domain be denoted as $\mathcal{D}_S =\left \{ \left ( x_S^i,y_S^i \right )  \right \}_{i=1}^{N^S}$, where $N^S$ is the number of labeled samples.
The target domain is represented as $\mathcal{D}_T=\left \{ \left ( x_{Tu}^j \right )  \right \}_{j=1}^{N^{Tu}}$, where $N^{Tu}$ denotes the number of unlabeled samples.
In line with the standard ADA setting \cite{xie2022active}, we introduce a labeling budget of size $N^B$, where $N^B \ll N^{Tu}$. 
This budget allows \system to select and label a small subset of the most informative samples from $\mathcal{D}_T$, forming a labeled target domain $\mathcal{D}_{Tl}=\left \{ \left ( x_{Tl}^j, y_{Tl}^j \right )  \right \}_{j=1}^{N^B}$.
We aim to train a model that adapts from the source domain $\mathcal{D}_S$ to the target domain $\mathcal{D}_T$, such that it generalizes well not only to the labeled samples but also to the remaining unlabeled data $\mathcal{D}_{Tu}$. Specifically, the target domain may exhibit agnostic domain shifts due to new positions, users, environments, or a combination thereof. By leveraging active learning, our approach minimizes labeling effort while maximizing adaptation performance, ensuring the model's robustness across various domains.

\subsection{EDL-based Activity Recognizer}\label{sec:dpn}
\begin{figure}[t]
    \centering
    \includegraphics[width=.38\textwidth]{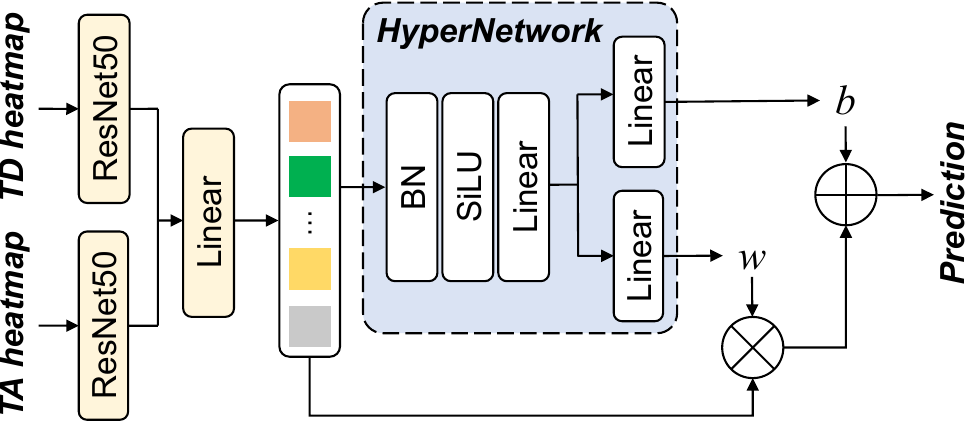}
    \caption{Architecture of the EAR network.}
    \label{fig:de_EAR}
    % \vspace{-0.15in}
\end{figure}

To enable reliable Human Activity Recognition (HAR) across diverse domains, we propose the Evidential Deep Learning-based Activity Recognizer (EAR).
Unlike traditional classifiers that produce only deterministic outputs, EAR leverages Evidential Deep Learning (EDL) \cite{sensoy2018evidential} to quantify prediction uncertainty. This capability is crucial for selecting informative samples, especially in new domains or under unseen conditions.
As shown in \Figure{de_EAR}, EAR comprises two main components: a feature extractor backbone and a HyperNetwork-based classifier. The backbone includes two independent ResNet50 networks \cite{he2016deep}, pretrained on ImageNet, to extract features from TD and TA heatmaps. These features are then fused via a linear layer in the latent space. The fused representation is passed to a HyperNetwork-based classifier \cite{zhang2024revisiting}, which predicts activity labels.

\textbf{Uncertainty Modeling with Dirichlet Prior}.
Inspired by \cite{sensoy2018evidential, xie2023dirichlet}, we place a Dirichlet prior on the predicted class probabilities of EAR. This prior forms the foundation for modeling both \textit{domain uncertainty} (due to domain shifts) and \textit{prediction uncertainty} (due to ambiguity in the data).
Following \cite{malinin2018predictive}, we define the class probability vector $\boldsymbol{\mu}$ (over class label $y$) for a given sample $\boldsymbol{x_i}$ as a Dirichlet distribution with concentration parameter $\boldsymbol{\alpha}$, where $\boldsymbol{\alpha} = \exp\left( O(x_i) \right)$ and $O(\cdot)$ denotes the output of EAR. $\boldsymbol{\mu}$ can be denoted as \Equation{mu} and the probability density function of Dirichlet distribution is defined as \Equation{Dirichlet}:
\begin{equation}
    \label{eq:mu}
    \boldsymbol{\mu} = [P(y=1), P(y=2), \dots, P(y=C)]^\top,
\end{equation}
\begin{small}
\begin{equation}
    \label{eq:Dirichlet}
    \operatorname{Dir}(\boldsymbol{\mu} \mid \boldsymbol{\alpha})=\frac{\Gamma\left(\boldsymbol{\alpha}_{0}\right)}{\prod_{c=1}^{C} \Gamma\left(\boldsymbol{\alpha}_{c}\right)} \prod_{c=1}^{C} \mu_{c}^{\boldsymbol{\alpha}_{c}-1}, \quad \boldsymbol{\alpha}_{c}>0, \boldsymbol{\alpha}_{0}=\sum_{c=1}^{C} \boldsymbol{\alpha}_{c},
\end{equation}
\end{small}
where $C$ denotes the number of activity classes and $\Gamma(\cdot)$ is Gamma function.

Based on this formulation, the posterior predictive distribution $\boldsymbol{\omega}$ is defined as:
% \begin{small}
\begin{equation}
    \label{eq:posterior}
    \mathrm{P}\left(\boldsymbol{\omega}_c \mid \boldsymbol{x}\right)=\int \mathrm{p}\left(\boldsymbol{\omega}_c \mid \boldsymbol{\mu}\right) \mathrm{p}\left(\boldsymbol{\mu} \mid \boldsymbol{x}\right) d \boldsymbol{\mu}=\operatorname{E}[\boldsymbol{\mu} \mid \boldsymbol{x}] =\frac{\boldsymbol{\alpha}_{c}}{\boldsymbol{\alpha}_{0}},
\end{equation}
% \end{small}
which provides both the predicted label and its associated uncertainty.

\textbf{Uncertainty vs Cognitive Domain Mismatch.}
This uncertainty serves as an indicator of cognitive domain mismatch. If the Dirichlet distribution over $\boldsymbol{\mu}$ is flat, it implies high uncertainty—suggesting that the sample lies outside the model's cognitive domain and likely originates from an unseen distribution \cite{malinin2018predictive}. Conversely, a sharp Dirichlet distribution indicates high confidence and familiarity. Therefore, uncertainty in $\boldsymbol{\mu}$ directly reflects the degree of domain shift, which is the central challenge in domain adaptation.

Through the use of a Dirichlet prior, EAR not only produces class probabilities but also provides a principled mechanism for measuring sample uncertainty. This makes it well-suited for real-world HAR scenarios, where data distributions are constantly shifting because of the agnostic domain shifts.

\subsection{R$\acute{e}$nyi Entropy-based Uncertainty Selector}\label{sec:uncertainty}

Effective active domain adaptation requires accurate identification of the most informative samples.
Existing methods primarily rely on Shannon entropy \cite{prabhu2021active}\cite{xie2023dirichlet}\cite{zhang2024revisiting} to measure uncertainty. We argue that R$\acute{e}$nyi entropy offers superior sensitivity and consistency across domains. 
% as it provides greater sensitivity to uncertainty variations and generates more consistent uncertainty across domains, making it more effective for domain adaptation.

R$\acute{e}$nyi entropy \cite{mayoral1998renyi} for a random variable $X$ is defined as:
\begin{equation}
    \label{eq:renyi_entropy}
    \begin{split}
    & \forall s > 0, s \neq 1\\
    &H_s(X) = \begin{cases}
        \frac{1}{1-s} \log \left( \sum_{c=1}^{C} p(X_c)^{s}\right), &  \text{if $X$ is discrete} \\
        \frac{1}{1-s} \log \int p(X)^{s} d X, & \text{if $X$ is continuous} 
         \end{cases}
    \end{split}
\end{equation}
where $p(\cdot)^s$ represents $s$-tilted probability distribution, also known as the escort distribution \cite{amari2016information}. 
Notably, as $s \rightarrow 1$, R$\acute{e}$nyi entropy converges to Shannon entropy. However, for $0<s<1$, it produces larger values, which could assign more weight to high-uncertainty (hard-to-classify) samples and guide the model to focus on these during training. This is particularly beneficial in domain adaptation, where performance depends on correctly handling such challenging cases.

\begin{figure*}[t]
    \centering
    \includegraphics[width=.75\textwidth]{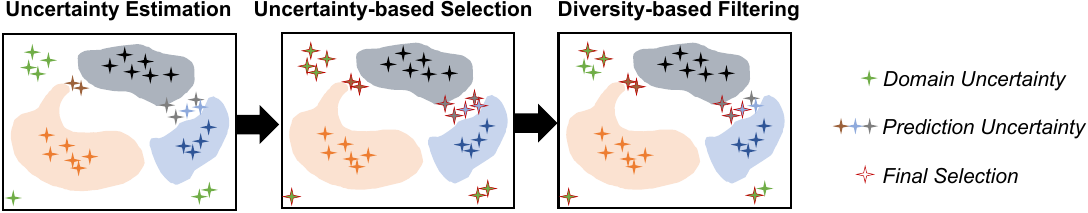}
    \caption{The procedure of uncertainty-based sample selection and filtering.}
    \label{fig:de_sampleSelection}
        % \vspace{-0.15in}
\end{figure*}

\textbf{Quantifying Domain and Prediction Uncertainty.}
We quantify two complementary uncertainties: domain uncertainty and prediction uncertainty.

Domain uncertainty captures how unfamiliar a sample is to the model, and is estimated via R$\acute{e}$nyi mutual information (MI) between the predicted label $\omega$ and the distribution $\boldsymbol{\mu}$. Since the uncertainty of $\boldsymbol{\mu}$ reflects domain shift, MI between $y$ and $\boldsymbol{\mu}$ effectively quantifies the class uncertainty induced by domain shift.
Given a sample $x_i$, we define EAR's domain uncertainty as:
% \begin{small}
\begin{equation}
    \label{eq:domain_uncertainty}
    \begin{split}
    \mathcal{U}_{\text {dom}}(x_i) &= I_s(\boldsymbol{\omega}_i \mid \boldsymbol{\mu_i}) = \frac{1}{1-s} \log \left( \sum_{c=1}^{C} (\frac{\boldsymbol{\alpha}_{ic}}{\boldsymbol{\alpha}_{i0}})^s \right) \\&- \frac{1}{1-s} \log \left( \sum_{c=1}^C \frac{\Gamma(\boldsymbol{\alpha}_{i0})\Gamma(\boldsymbol{\alpha}_{ic}+s)}{\Gamma(\boldsymbol{\alpha}_{ic})\Gamma(\boldsymbol{\alpha}_{i0}+s)} \right),
    \end{split}
\end{equation} 
% \end{small}
where we adopt Hayashi's R$\acute{e}$nyi MI \cite{hayashi2011exponential} for computational efficiency (see Appendix A for detailed derivations).

On the other hand, prediction uncertainty reflects ambiguity within the cognitive domain and is estimated using R$\acute{e}$nyi conditional entropy. 
The conditional entropy of $y$ given $\boldsymbol{\mu}$ quantifies the remaining uncertainty in the class prediction when $\boldsymbol{\mu}$ is known. Even if a sample lies within the model's cognitive domain, it may still exhibit high uncertainty due to its proximity to class boundaries—this ambiguity stems from the sample's intrinsic characteristics, such as its similarity to another activity, rather than from domain shift.
It is defined as (the detailed derivation is provided in Appendix A):

\begin{small}
\begin{equation}
    \label{eq:predictive_uncertainty}
    \mathcal{U}_{\text {pred}}(x_i) = H_s(\boldsymbol{\omega}_i \mid \boldsymbol{\mu_i}) = \frac{1}{1-s} \log \left( \sum_{c=1}^C \frac{\Gamma(\boldsymbol{\alpha}_{i0})\Gamma(\boldsymbol{\alpha}_{ic}+s)}{\Gamma(\boldsymbol{\alpha}_{ic})\Gamma(\boldsymbol{\alpha}_{i0}+s)} \right).
\end{equation}
\end{small}

Together, these yield the total uncertainty:
\begin{equation}
    \label{eq:total_uncertainty}
    \mathcal{U}_{\text {total}}(x_i) = \lambda_{\text {dom}} \mathcal{U}_{\text {dom}}(x_i) + \lambda_{\text {pred}} \mathcal{U}_{\text {pred}}(x_i).
\end{equation}
Since domain shift is the primary focus in adaptation, the weighting factor $\lambda_{\text {dom}}$ for $\mathcal{U}_{\text {dom}}(x_i)$ is set significantly higher than $\lambda_{\text {pre}}$ for $\mathcal{U}_{\text {pred}}(x_i)$. This combined uncertainty effectively captures the most informative samples, regardless of the target domain being adapted.

\textbf{Sample Selection Strategy.} This total uncertainty guides our sample selection strategy for labeling. Following \cite{zhang2024revisiting}, we adopt a two-step process:
i) \textit{Uncertainty-based Selection}. In round $i$, we identify the top $(i+1)N_i^B$ samples with the highest total uncertainty as candidates for labeling, where $N_i^B$ is the labeling budget of this round.
ii) \textit{Diversity-aware Filtering}. From these candidates, we further select $N_i^B$ samples that maximize $d(1+u)$, where $d$ is the normalized average latent feature diversity (based on dot-product distance) and $u$ is the normalized uncertainty. This ensures that selected samples are both uncertain and diverse, enhancing adaptation effectiveness.

\subsection{Model Refiner}
To further improve the domain adaptation performance of EAR, we introduce two model refinement techniques: pseudo labeling and contrastive learning.

\subsubsection{Pseudo Labeling}
By R$\acute{e}$nyi Entropy-based uncertainty selector, we construct $\mathcal{D}_{Tl}$. Labeled data in $\mathcal{D}_{Tl}$ provide direct supervision, but we consider the unlabeled data $\mathcal{D}_{Tu}$ in $\mathcal{D}_T$ also contain valuable target-domain information that can further improve model adaptation. A key observation is that samples from the same category tend to cluster together in a latent space, while samples from different classes remain distant. Accordingly, if two samples are sufficiently close in this space, they likely belong to the same category.

Building on this insight, we propose a new KNN-based similarity estimation method to assign pseudo labels to unlabeled samples in each selection round. Given a sample $x_i$, we define its similarity to class $c$ as $S(i, c)$, computed using the distance metric $d(i, j)$:

% \begin{small}
\begin{equation}
    \label{eq:knn_distance}
    \begin{gathered}
        d(i, j)=1-\frac{\Omega(x_i)}{\|\Omega(x_i)\|} \cdot \frac{\Omega(x_j)}{\|\Omega(x_j)\|}, \quad \forall x_i, x_j \in\left\{\mathcal{D}_S \cup \mathcal{D}_T\right\}\\
    \end{gathered}
\vspace{-0.1in}
\end{equation}
% \end{small}

\begin{footnotesize}
\begin{equation}
    \label{eq:similarity_score}
    % \begin{aligned}
    \begin{split}
    S(i, c) &= \sum_{x_j \in \mathcal{D}_S^c} \mathbb{I} \left\{ x_i \in \mathcal{K}_j \wedge x_j \in \mathcal{K}_i \right\} \cdot \mathbb{I} \left\{ d(i, j) < \overline{d}(i, j) \right\} \\ &+ 2 \cdot \sum_{x_j \in \mathcal{D}_{Tl}^c} \mathbb{I} \left\{ x_i \in \mathcal{K}_j \wedge x_j \in \mathcal{K}_i \right\} \cdot \mathbb{I} \left\{ d(i, j) < \overline{d}(i, j) \right\},\\
    % &\overline{d}(i, j) = \begin{cases}
    %     d(i, j)<\frac{1}{\left| \mathcal{D}_S \right|} \sum_{x_p \in \mathcal{D}_S} d(i, p)-3 * \sqrt{\frac{1}{\left| \mathcal{D}_S \right|} \sum_{x_p \in \mathcal{D}_S}\left(d(i, p)-\frac{1}{\left| \mathcal{D}_S \right|} \sum_{x_p \in \mathcal{D}_S} d(i, p)\right)^{2}}, &  x_j \in \mathcal{D}_S \\
    %     d(i, j)<\frac{1}{\left| \mathcal{D}_T \right|} \sum_{x_p \in \mathcal{D}_T} d(i, p)-3 * \sqrt{\frac{1}{\left| \mathcal{D}_T \right|} \sum_{x_p \in \mathcal{D}_T}\left(d(i, p)-\frac{1}{\left| \mathcal{D}_T \right|} \sum_{x_p \in \mathcal{D}_T} d(i, p)\right)^{2}}, &  x_j \in \mathcal{D}_T    
    %      \end{cases}\\
     \end{split}
    % \end{aligned}
\end{equation}
\end{footnotesize}
where $\Omega(\cdot)$ denotes the backbone of EAR. The similarity $S(i, c)$ is then calculated based on the K-nearest neighbors $\mathcal{K}_i$ and $\mathcal{K}_j$ of $x_i$ and $x_j$, respectively, in the latent space.
The distance upper bound $\overline{d}(i, j)$ is determined using the 3-sigma principle (refer to Appendix B for the definition), ensuring that we identify close samples with 99.7\% confidence.
$\mathbb{I}{\cdot}$ is an indicator function.
The sets $\mathcal{D}_S^c$ and $\mathcal{D}_T^c$ represent the collections of samples belonging to class $c$ in the source and target domains, respectively.

Finally, the pseudo label $\hat{y_i}$ of $x_i$ is inferred as \Equation{pseudo_label}:
\begin{equation}
    \label{eq:pseudo_label}
    \hat{y_i} = \left\{ q \bigg| S(i,q) > 0 \wedge S(i,q) > \frac{1}{C} \sum_{c=1}^{C} S(i, c) \right\}.
\end{equation}
Notably, unlike traditional pseudo labeling approaches that assign a single label per sample, our method produces a set of labels, named Pseudo Label Set (PLS), which can be empty, contain one label, or multiple labels. This adaptive labeling strategy provides a more reliable estimate, especially for ambiguous or unfamiliar samples, reducing the risk of erroneous pseudo-label assignments, which is a common issue in standard pseudo-labeling techniques.

\subsubsection{Contrastive Learning}

We further incorporate contrastive learning to enhance class-level alignment in the latent space. This approach encourages samples from the same activity category to cluster more tightly while pushing apart those from different categories. By refining the latent representation in this manner, contrastive learning not only improves classification accuracy but also strengthens class consistency, thereby increasing the reliability of pseudo-label inference.

The contrastive loss function is defined as:

\begin{scriptsize}
\begin{equation}\label{eq:contrastive_learning}
\begin{split}
    \mathcal{L}_c &= \frac{1}{\left| \mathcal{D}_S \cup \mathcal{D}_{Tl} \right|} \sum_{x_i \in \left\{ \mathcal{D}_S \cup \mathcal{D}_{Tl} \right\}} \left( \frac{\sum_{x_j \in \widetilde{\mathcal{D}_S^{y_i}}}d(i, j)}{\sum_{x_j \in \mathcal{D}_S}d(i, j)}
    + \frac{\sum_{x_j \in \widetilde{\mathcal{D}_{Tl}^{y_i}}}d(i, j)}{\sum_{x_j \in \mathcal{D}_{Tl}}d(i, j)} \right) \\&+ \frac{1}{\left| \mathcal{D}_{Tl} \right|} \sum_{x_i \in \mathcal{D}_{Tl}} \left( \frac{1}{\left| \widetilde{\mathcal{D}_S^{y_i}} \right|} \sum_{x_j \in \widetilde{\mathcal{D}_S^{y_i}}}d(i, j) \right),
\end{split}
\end{equation}
\end{scriptsize}
where $\widetilde{\mathcal{D}_S^{y_i}}$ and $\widetilde{\mathcal{D}_{Tl}^{y_i}}$ are randomly sampled subsets of $\mathcal{D}_S^{y_i}$ and $\mathcal{D}_{Tl}^{y_i}$, which respectively contain data of class $y_i$ from the source domain $\mathcal{D}_S$ and the labeled target domain samples $\mathcal{D}_{Tl}$ in current selection round. 
The term $d(i,j)$ represents the distance metric used to measure feature similarity in the latent space (\Equation{knn_distance}).
%
% By leveraging contrastive learning, our model learns a better-structured latent representation, making it more robust for domain adaptation. This, in turn, reduces ambiguity in pseudo labeling and enhances the model's generalization ability across domains.

\subsection{Loss Function}
\textbf{Evidential Loss.}
To effectively extract meaningful information from both labeled and pseudo-labeled samples, we employ the negative log-likelihood (NLL) loss $\mathcal{L}_{nll}(x_i, y_i)$ to train the EAR:
% \begin{small}
\begin{equation}
    \label{eq:nll}
    \mathcal{L}_{nll}(x_i, y_i) = - \sum_{c=1}^{C} \beta_{ij} \left( \log (\sum_{k=1}^{C}\boldsymbol{\alpha}_{ik}) - \log(\boldsymbol{\alpha}_{ic}) \right),
\end{equation}
% \end{small}
where $\beta_{ij}$ is the $j_{th}$ element of the one-hot vector corresponding to $y_i$.
However, $\mathcal{L}_{nll}$ loss alone can bring invalid evidence into other class predictions, potentially degrading model performance. To mitigate this, we also incorporate KL divergence loss $\mathcal{L}_{kl}(x_i, y_i)$, which regulates the training process by reducing the influence of unreliable evidence \cite{sensoy2018evidential}: 
\begin{equation}
    \label{eq:kl_divergence}
    \mathcal{L}_{kl}(x_i, y_i) = KL\left[\operatorname{Dir}\left(\boldsymbol{p} \mid \hat{\boldsymbol{\alpha}_i}\right) \mid \operatorname{Dir}(\boldsymbol{p} \mid \boldsymbol{1})\right],
\end{equation}
where $\hat{\boldsymbol{\alpha}_i} = \beta_{iy_i}+\left(1-\beta_{iy_i}\right) \cdot \boldsymbol{\alpha}_i$. Here, $\hat{\boldsymbol{\alpha}_i}$ denotes the misleading evidence from $x_i$ that could negatively impact class separability \cite{sensoy2018evidential}. The term $\operatorname{Dir}(\boldsymbol{\mu}_i \mid \boldsymbol{1})$ corresponds to a uniform Dirichlet distribution, which serves as a reference. By minimizing $\mathcal{L}_{kl}(x_i, y_i)$, we can reduce the influence of incorrect evidence, improving the model's robustness in uncertain scenarios.
The total evidential loss for each sample can be expressed as:

\begin{footnotesize}
\begin{equation}\label{eq:edl_loss}
    \begin{split}
    \mathcal{L}_{edl} &= \frac{1}{\left| \mathcal{D}_S \cup \mathcal{D}_{Tl} \right|} \sum_{x_i \in \left\{ \mathcal{D}_S \cup \mathcal{D}_{Tl} \right\}} \left( \mathcal{L}_{nll}(x_i, y_i) + \mathcal{L}_{kl}(x_i, y_i) \right) \\ 
    &+ \frac{1}{\left| \mathcal{D}_{Tu} \right|} \sum_{x_i \in \mathcal{D}_{Tu}, \hat{y_i} \neq \emptyset } \left( \frac{1}{\left| \hat{y_i} \right|} \sum_{\widetilde{y_i}\in \hat{y_i}} (\mathcal{L}_{nll}(x_i, \widetilde{y_i}) + \mathcal{L}_{kl}(x_i, \widetilde{y_i})) \right) 
    \end{split}
\end{equation}
\end{footnotesize}

\textbf{Uncertainty-based Alignment Loss.}
Beyond evidential learning, domain alignment also plays a crucial role. To reduce the domain gap, we introduce an uncertainty-based alignment loss $\mathcal{L}_{u}$, which accounts for both domain uncertainty $\mathcal{U}_{\text {dom}}(x_i)$ (\Equation{domain_uncertainty}) and prediction uncertainty $\mathcal{U}_{\text {pred}}(x_i)$ (\Equation{predictive_uncertainty}) on $\mathcal{D}_{Tu}$.
Minimizing these uncertainties help mitigate the domain shift between the source and target domain.
The uncertainty loss is formulated as:
% \begin{small}
\begin{equation}
    \label{eq:uncertainty_loss}
    \mathcal{L}_{u} = \frac{1}{\left| \mathcal{D}_{Tu} \right|} \sum_{x_i \in \mathcal{D}_{Tu}} \left( \lambda_{\text {dom}} \mathcal{U}_{\text {dom}}(x_i) + \lambda_{\text {pred}} \mathcal{U}_{\text {pred}}(x_i) \right)
\end{equation}
% \end{small}

\textbf{R$\acute{e}$nyi Entropy Regularization Loss.}
Additionally, to enhance the robustness of the system, we make the R$\acute{e}$nyi order $s$ in \Equation{renyi_entropy} learnable. However, as indicated by \Equation{renyi_entropy}, R$\acute{e}$nyi entropy of a sample decreases as $s$ increases. If $s$ continues to grow without constraint during training, it may exceed 1, which would reduce the sensitivity of R$\acute{e}$nyi entropy to hard samples. To address this, we introduce an additional regularization term, \ie, $\mathcal{L}_s = (s-\frac{1}{2})^2$. This ensures that $ 0 < s < 1 $ throughout the training process, maintaining the effectiveness of entropy-based uncertainty estimation.

\textbf{Summary.} The final training objective combines the aforementioned losses, defined as follows:
\begin{equation}
    \label{total loss}
    \mathcal{L}_{total} = \mathcal{L}_{edl} + \mathcal{L}_{u} + \lambda_c\mathcal{L}_c + \mathcal{L}_s
\end{equation}
where $\lambda_c$ is a hyperparameter.
This formulation ensures that both labeled and pseudo-labeled data contribute effectively to model training. 
% By leveraging $\mathcal{L}_{edl}$ for evidence extraction, $\mathcal{L}_{u}$ for domain alignment, $\mathcal{L}_c$ for feature alignment, and $\mathcal{L}_s$ for regularization, our approach enhances the model's adaptability and generalization across domains.
\section{Implementation}

\subsection{Experimental Setups}

\subsubsection{Data Collection}

\begin{figure}[t]
    \centering
    \includegraphics[width=.47\textwidth]{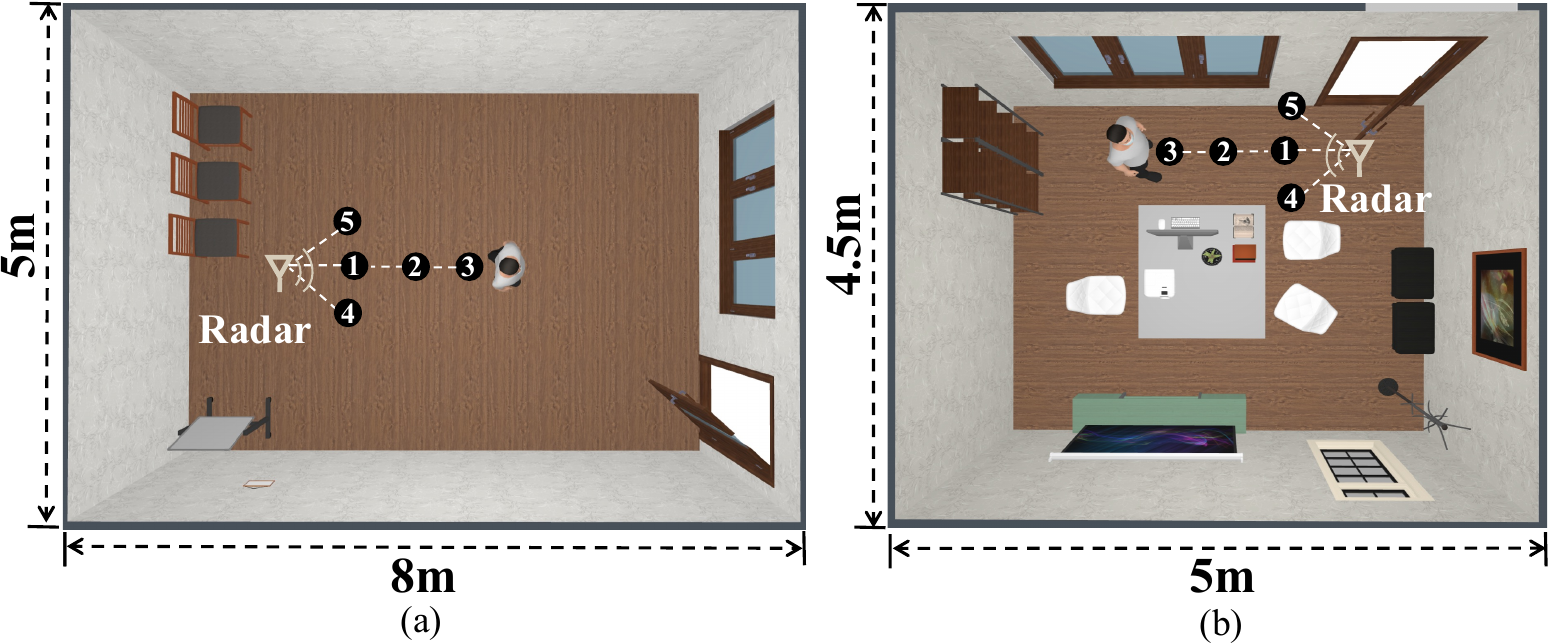}
    \vspace{-0.03in}
    \caption{Illustration of experimental scenes.}
    \label{fig:impl_scene}
    \vspace{-0.1in}
\end{figure}

We use a commercial mmWave radar (TI IWR1443BOOST) to collect FMCW sensing data. The radar transmits 20 frames per second, with each frame consisting of 120 chirps. The frequency range of the chirp spans from 77GHz to 81GHz. 
As illustrated in \Figure{impl_scene}, experiments are conducted in two indoor environments with varying room sizes and furniture layouts. Each environment includes five distinct positions. The radar is positioned at a height of approximately 1.2m, with the volunteer standing facing the radar.

\subsubsection{Dataset}
For data collection, we invite 8 volunteers to perform 12 activities across 5 positions in 2 different environments. 
The activities include: 1) Pushing, 2) Pulling, 3) Sliding right, 4) Sliding left, 5) Clapping, 6) Crossing arms, 7) Drawing a zigzag, 8) Drawing an M shape, 9) Drawing a circle clockwise, 10) Drawing a circle anticlockwise, 11) Waving hands, 12) Random action or no action. These actions involve both single- and double-arm motions.
Each volunteer performs each activity 10 times at every position, resulting in a total of 9600 samples, with an equal number of samples collected under each condition. 
Our experiment is conducted with approval from the Institutional Review Board (IRB).

\subsection{DNN Implementation Details}

The EAR model is trained using the SGD optimizer \cite{bottou2010large} with a cosine annealing scheduler \cite{loshchilov2016sgdr} on an NVIDIA GeForce RTX 3090. The initial learning rate for the extractors and $s$ (in \Equation{uncertainty_loss}) is $5 \times e^{-4}$, while other parameters use a learning rate of $5 \times e^{-3}$. 
Besides, we configure the following hyperparameters: batch size = 16, $\lambda_{\text {dom}}$ = 7, $\lambda_{\text {pred}}$ = 0.5, $\lambda_c$ = 1, $k = 5$. 
Following \cite{zhang2024revisiting}, we select $N^B = 5\% N^{Tu}$ target samples across five selection rounds, with $N_i^B = 1\% N^{Tu}$ for $i = 1, 2, \dots, 5$. 
Our model contains 0.05G parameters and requires 7.70GFLOPs per inference. The average inference latency is 12ms on the RTX 3090.

\section{Evaluation}

In this section, we present the experimental results of \system across various aspects.

\textbf{Metrics.} 
The goal of \system is to transfer the HAR model from the source domain to the target domain. 
To assess the performance, we use standard metrics for classification tasks: \textit{accuracy}, \textit{precision}, \textit{recall}, and \textit{F1-score}.

% \textbf{Baselines.} We compare \system with five baselines:

% $\bullet$ mTransSee \cite{liu2022mtranssee}: A fine-tuning-based transfer learning method that adapts mmWave-based gesture recognition systems to new environments.

% $\bullet$ EI \cite{jiang2018towards}: A wireless HAR framework that effectively removes environment- and subject-specific information from activity data using adversarial learning.

% $\bullet$ RoMF \cite{zhang2024few}: A meta-learning-based framework for wireless HAR that decouples the dependence on sensing conditions, enabling accurate activity recognition across new environments, users, and activity classes without the need for model fine-tuning.

% $\bullet$ EADA \cite{xie2022active}: An energy-based active domain adaptation method that incorporates both domain characteristics and free energy-based instance uncertainty for sample selection.

% $\bullet$ MADA \cite{zhang2024revisiting}: A state-of-the-art active learning-based multi-source domain adaptation method that uses EDL to identify informative target samples.

\begin{figure}[t]
\centering
\begin{minipage}[t]{\linewidth}
\centering
\subfigure[S1-T1\&2-U3\&4]
{\includegraphics[width=.45\textwidth]{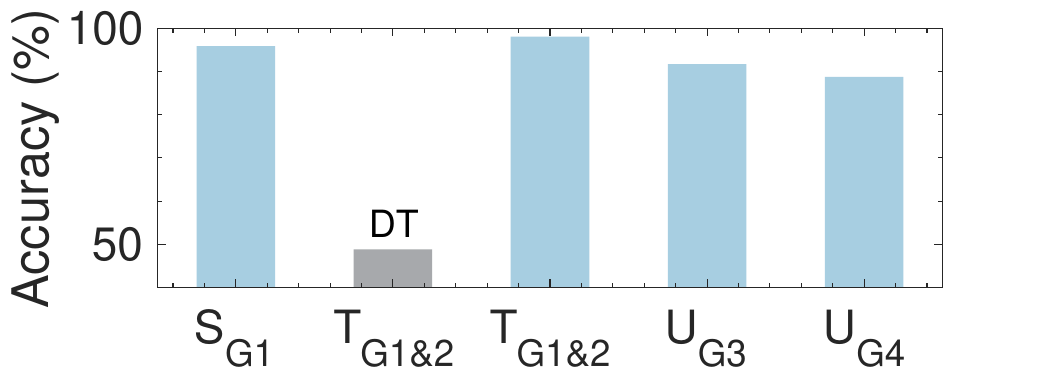}}
% \hspace{0.1in}
\subfigure[S2-T2\&3-U1\&4]
{\includegraphics[width=.45\textwidth]{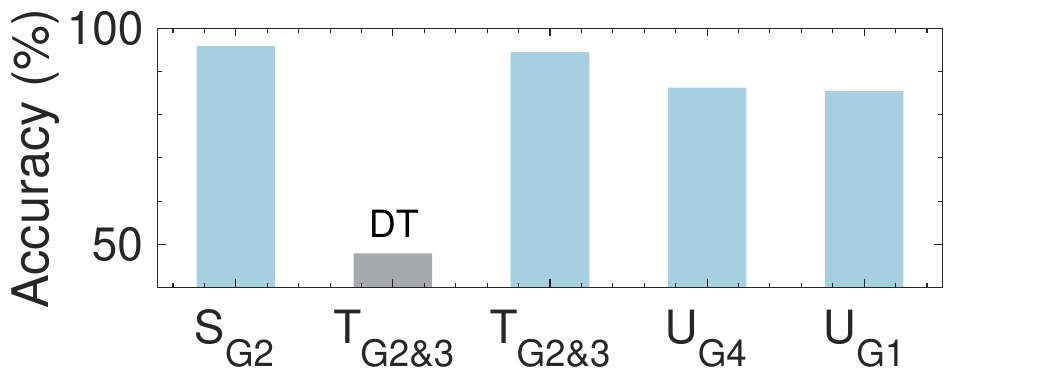}}
\subfigure[S3-T3\&4-U1\&2]
{\includegraphics[width=.45\textwidth]{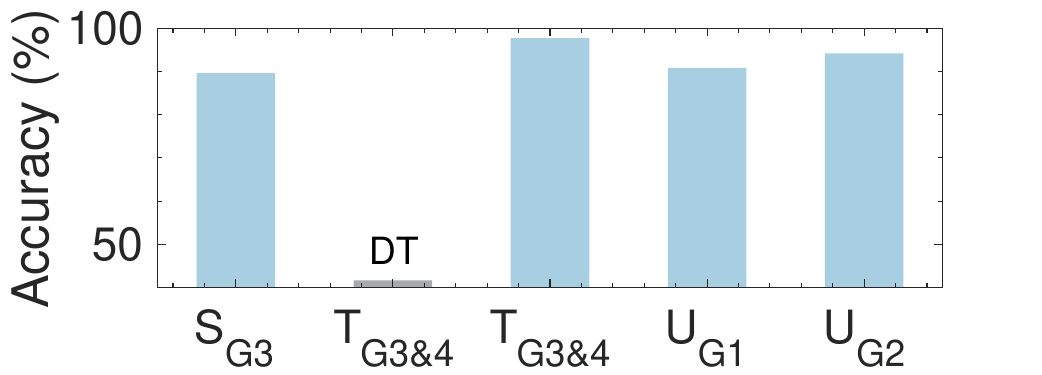}}
% \hspace{0.1in}
\subfigure[S4-T4\&1-U2\&3]
{\includegraphics[width=.45\textwidth]{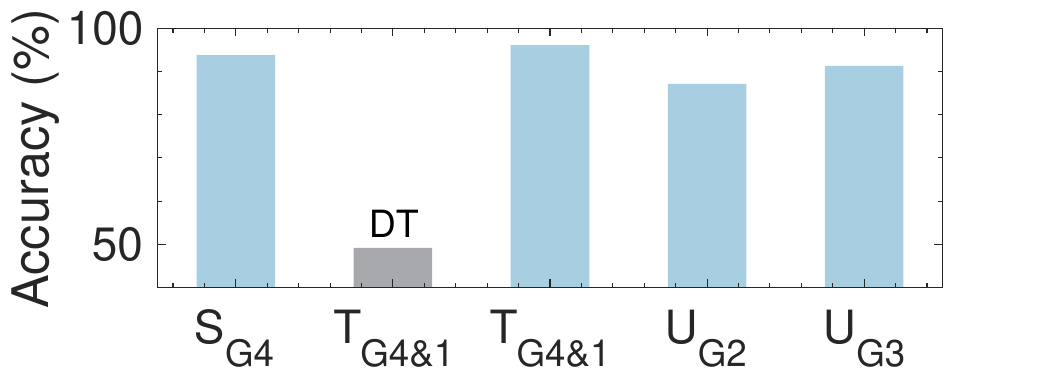}}
% \vspace{-0.03in}
\caption{Comprehensive domain adaptation experimental results when regarding different groups of users as the target domain. S: Source domain. T: Target Domain. U: Unseen Domain.}
\label{fig:exp_overall}
\end{minipage}
\vspace{-0.2in}
\end{figure}

\subsection{Overall Performance}

We evaluate the overall performance of \system and compare it with multiple baseline methods. Scene \#1 (\Figure{impl_scene}(a)) serves as the default environment, while Scene \#2 is used as an unseen environment to assess \system's generalization capability. Data from eight users is divided into four groups: Group \#1 (users \#1, \#2), Group \#2 (users \#3, \#4), Group \#3 (users \#5, \#6), and Group \#4 (users \#7, \#8). In the default setting, we select Pos \#2 from two users in one group as the source domain, while their remaining positions and another group serve as the target domain. The remaining two groups are treated as the unseen domains. 
% \textcolor{red}{The source domain only contains data of two users from one position, which is a demanding experimental setup, to test the ability of \system under harsh conditions}

To demonstrate \system's effectiveness, we designate each group as the source domain in separate experiments. \Figure{exp_overall} presents the HAR accuracy for the source, target, and unseen domains. The gray bars represent direct transfer (DT) results, where the model trained on the source domain is applied directly to the target domain without adaptation. The significant drop in accuracy highlights the impact of domain shift. However, with \system's domain adaptation techniques, performance improves, consistently achieving over 90\% accuracy across both target and unseen domains. For a more detailed analysis, we use `S1-T1\&2-U3\&4' as the default setting in the following.

\subsubsection{Performance Across Different Activities}

\begin{figure}[t]
\begin{minipage}[h]{0.45\linewidth}
\centering
    \includegraphics[width=\textwidth]{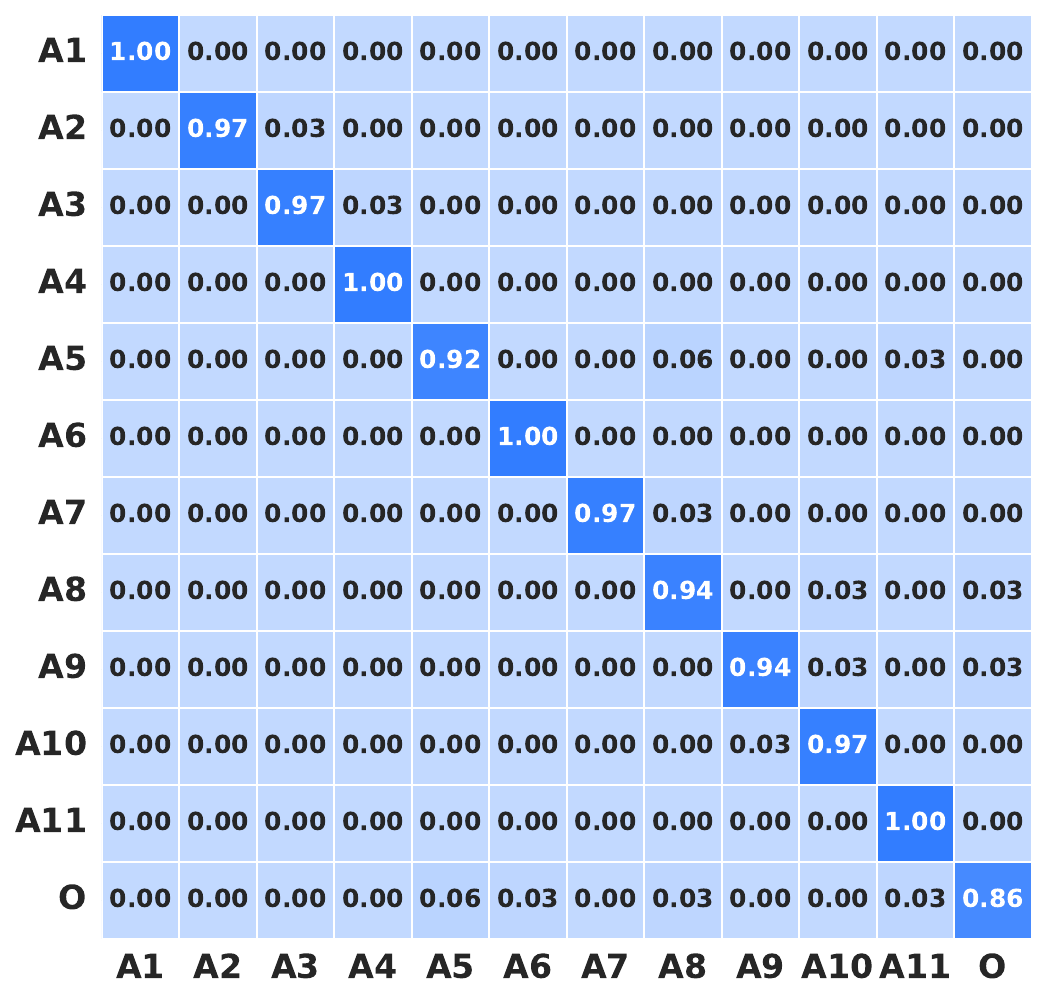}
    \caption{Confusion Matrix.} % when Group \#1 is the source domain and Group \#2 is target domain
    \label{fig:exp_ConfusionMatrix}
\end{minipage}
\begin{minipage}[h]{0.45\linewidth}
\centering
    \includegraphics[width=\textwidth]{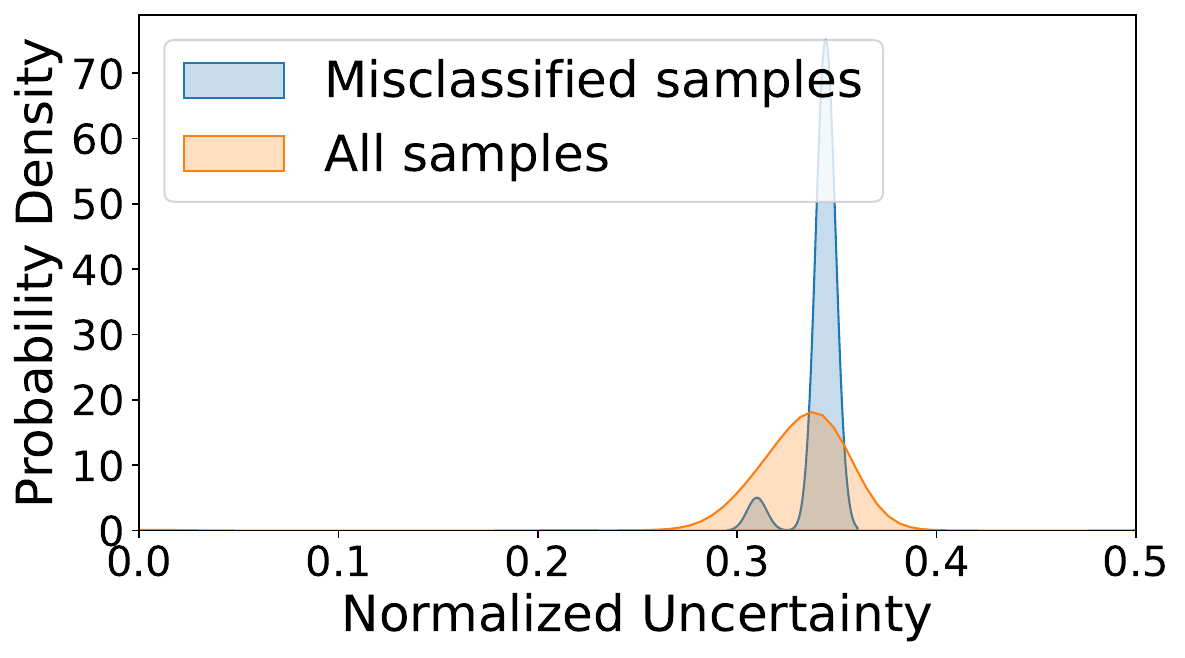}
    \caption{Kernel Density Estimation (KDE) of uncertainty scores for misclassified samples compared to all samples.} % when Group \#1 is the source domain and Group \#2 is target domain
    \label{fig:exp_kde_uncertainties}
\end{minipage}
\end{figure}

To assess class-wise recognition performance, we present the confusion matrix in \Figure{exp_ConfusionMatrix}. All activities achieve over 90\% accuracy in the target domain, demonstrating that \system effectively preserves action-level discrimination despite domain shifts.
Further analysis in \Figure{exp_kde_uncertainties} shows that most misclassified samples are associated with high uncertainty scores. This validates \system's uncertainty estimation, confirming its ability to identify ambiguous or hard-to-classify instances.
Interestingly, many of the selected labeled samples are from activity pairs that are inherently difficult to distinguish due to subtle differences—for example, A7 (Drawing a zigzag) vs. A8 (Drawing an M shape), and A9 (Drawing a circle clockwise) vs. A10 (Drawing a circle anticlockwise). These four activities alone account for over 47.8\% of the 5\% labeling budget, reinforcing the idea that \system prioritizes labeling samples from similar and easily confused categories to enhance overall performance.

% \begin{figure*}[h]
% \begin{minipage}[h]{0.32\linewidth}
% \centering
%     \includegraphics[width=\textwidth]{pic/exp_ConfusionMatrix}
%     \caption{Confusion Matrix.} % when Group \#1 is the source domain and Group \#2 is target domain
%     \label{fig:exp_ConfusionMatrix}
% \end{minipage}
% \hspace{0.002in}
% \begin{minipage}[h]{0.33\linewidth}
% \centering
% \includegraphics[width=\textwidth]{pic/exp_overall_diffUsers}
% \caption{Performance across users.}
% \label{fig:exp_overall_diffUsers}
% \end{minipage}
% % \hspace{0.01in}
% \begin{minipage}[h]{0.33\linewidth}
% \centering
% \includegraphics[width=\textwidth]{pic/exp_overall_diffDomains}
% \caption{Performance across positions.}
% \label{fig:exp_overall_diffDomains}
% \end{minipage}
% \vspace{-0.1in}
% \end{figure*}

\begin{figure}[t]
\begin{minipage}[h]{0.45\linewidth}
\centering
\includegraphics[width=\textwidth]{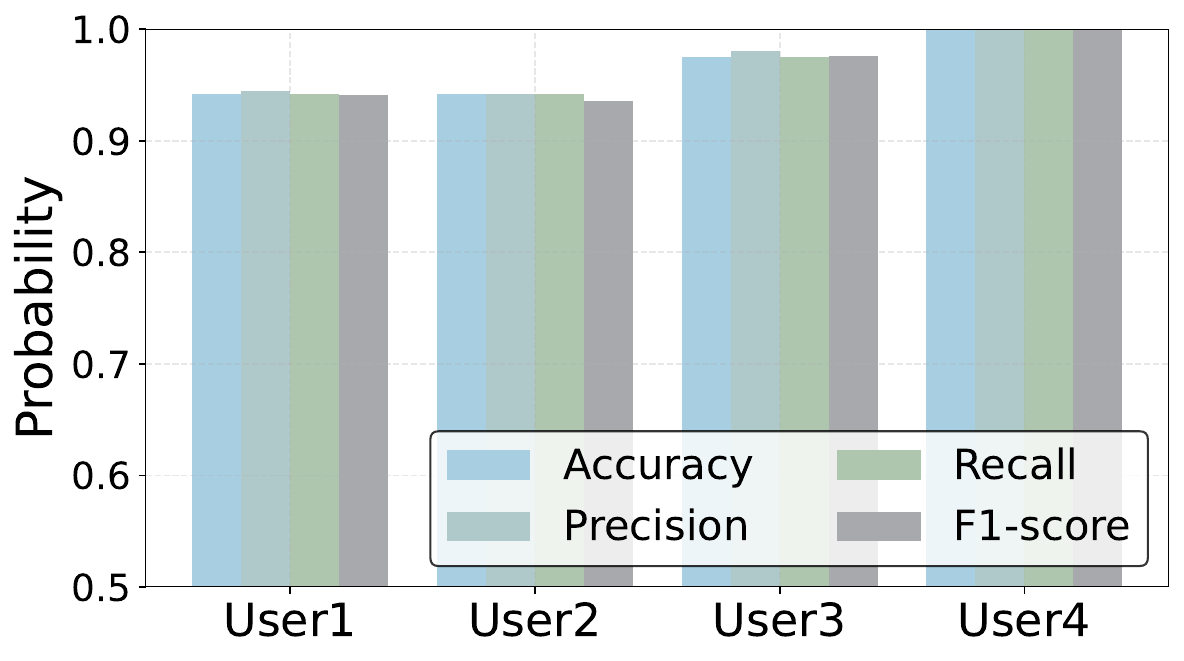}
\caption{Performance vs users.}
\label{fig:exp_overall_diffUsers}
\end{minipage}
% \hspace{0.01in}
\begin{minipage}[h]{0.45\linewidth}
\centering
\includegraphics[width=\textwidth]{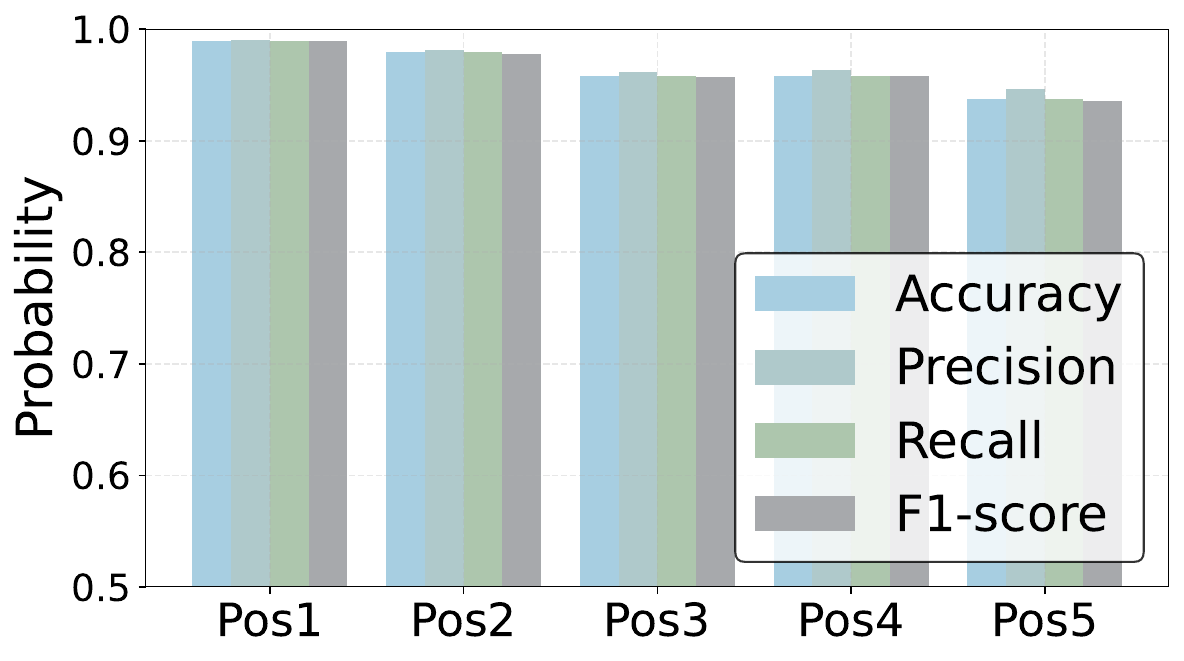}
\caption{Performance vs positions.}
\label{fig:exp_overall_diffDomains}
\end{minipage}
% \begin{minipage}[h]{0.245\linewidth}
% \centering
%     \includegraphics[width=.8\textwidth]{pic/exp_location_of_misclassified_samples}
%     \caption{Collect locations of misclassified samples.} % when Group \#1 is the source domain and Group \#2 is target domain
%     \label{fig:exp_location_of_misclassified_samples}
% \end{minipage}
\vspace{-0.1in}
\end{figure}

% \begin{figure}[t]
% % \hspace{0.002in}
% \begin{minipage}[h]{0.45\linewidth}
% \centering
%     \includegraphics[width=.85\textwidth]{pic/exp_selected_samples_pos}
%     \caption{Collect location of selected samples.} % when Group \#1 is the source domain and Group \#2 is target domain
%     \label{fig:exp_selected_samples_pos}
% \end{minipage}
% \begin{minipage}[h]{0.45\linewidth}
% \centering
%     \includegraphics[width=.85\textwidth]{pic/exp_selected_samples_activity}
%     \caption{Activity of selected samples.} % when Group \#1 is the source domain and Group \#2 is target domain
%     \label{fig:exp_selected_samples_activity}
% \end{minipage}
% % \vspace{-0.1in}
% \end{figure}

\subsubsection{Performance Across Different Users}
Next, we evaluate HAR performance for individual users in the target domain (\Figure{exp_overall_diffUsers}). Generally, each user's physical characteristics (\eg, height, weight) and habits influence the reflected mmWave signals, causing variations in mmWave signatures (TD and TA heatmaps). The results show that \system's active learning strategy effectively selects the most informative samples for annotation, enabling the HAR model to adapt across users with $>$95\% accuracy.

\subsubsection{Performance Across Different Positions}
We next assess HAR performance across various positions within Scene \#1. Among them, only Pos \#2 from Users \#1 and \#2 is part of the source domain; all other positions represent the target domain. This setup allows us to assess \system's ability to adapt to different user-to-radar distances and angles.
As shown in \Figure{exp_overall_diffDomains}, performance at Pos \#4 and \#5 is comparatively moderate. To understand this, we analyze the distribution of selected labeled samples across positions. The proportions for Pos \#1 to Pos \#5 are 10.8\%, 6.2\%, 26.2\%, 27.7\%, and 29.2\%, respectively. This pattern reflects the nature of mmWave signal propagation—greater distances and wider angles result in stronger signal attenuation, necessitating more supervision for reliable adaptation. In particular, larger angles significantly distort both velocity and angle features, contributing to the relatively lower performance at Pos \#4 and \#5.
Despite these challenges, \system successfully identifies and labels the most informative samples, enabling it to maintain over 90\% accuracy across all positions.

% \subsubsection{Composition of Selected Samples}
% \textcolor{red}{\Figure{exp_selected_samples_pos} shows that most of selected samples are collected at Pos3, Pos4 and Pos5, which are challenging locations for HAR because of large distance and angle. \Figure{exp_selected_samples_activity} shows that selected samples focus on similar activities that are difficult to distinguish due to small differences, such as A7(Drawing a zigzag) and A8(Drawing an M shape), A9(Drawing a circle clockwise) and A10(Drawing a circle anticlockwise). Difficult positions and activities indeed lead to high uncertainty in new domain so that samples belonging to them are selected more. These results suggest that \system can adaptively select more samples belonging to difficult activities collected at difficult positions while containing sample diversity.}

% \begin{small}
\begin{table}[h]
\centering
  \caption{Overall performance (\%) and comparisons. RE: R$\acute{e}$nyi Entropy. PL: Pseudo Labeling. PLS: Pseudo Label Set. CL: Contrast Learning.}
  \setlength{\tabcolsep}{1.5pt}
 % \vspace{-0.1in}
  \begin{tabular}{ccccl}
    \toprule
     & Accuracy   &  Precision & Recall  & F1-score \\
    \hline
    \textbf{\system~(Ours)}    & \bf{96.30} & \bf{96.33}  & \bf{96.30}  & \bf{96.28}\\
    Source Domain G1    & 97.92 & 98.33 & 97.92 & 97.88        \\
    Unseen Domain G3    &  91.25 & 91.91 & 91.25 & 90.90    \\
    Unseen Domain G4    &  92.50 & 92.70 & 92.50 & 92.49    \\
    w/o RE   &  93.75  &  93.89 &  93.75  &  93.76  \\
    w/o PL   &  81.48 &86.67  &81.48  &82.91\\
    w/o PLS   &  65.97 &81.52  &65.97  &69.43\\
    w/o CL   &  91.90 &92.01  &91.90  &91.78\\
       \hline
    Direct Transfer & 43.52 & 47.76 & 43.52 &42.86 \\
    mTransSee~\cite{liu2022mtranssee}     &  45.45  & 61.47 & 45.45 & 47.73\\
    EI~\cite{jiang2018towards}  &  79.17 & 85.14 & 79.17 & 78.22\\
    RoMF~\cite{zhang2024few}  & 31.57 & 51.13 & 31.57 & 31.28\\
    EADA~\cite{xie2022active}    & \underline{85.19} & \underline{86.89} & \underline{85.19} & \underline{84.51}\\
    MADA~\cite{zhang2024revisiting}   & 75.69 & 83.94 & 75.69 & 76.16\\
  \bottomrule
\end{tabular}
\label{tab:exp_overall_comparison}
% \vspace{-0.1in}
\end{table}
% \end{small}

\subsection{Comparison with Baseline Methods}
We compare \system with five representative baseline methods. \Table{exp_overall_comparison} shows that our approach achieves state-of-the-art performance, with accuracy, precision, recall, and F1-score all reaching 96.28\%–96.33\%. Furthermore, \system maintains strong performance in unseen domains, achieving over 90\% accuracy, demonstrating its generalization capability.
We also have several observations emerge from the comparison:
(1) mTransSee \cite{liu2022mtranssee} pre-trains a HAR model on the source domain and fine-tunes it using 5\% of labeled target data. When tested on our dataset, it achieves 45.45\% accuracy, 61.47\% precision, 45.45\% recall, and 47.73\% F1-score, significantly lower than \system. This gap suggests that \system's active sample selection is more effective than the random selection typically used in fine-tuning.
(ii) EI \cite{jiang2018towards}, an adversarial learning method, eliminates domain-specific features using a domain discriminator. However, its performance remains moderate. Since our source domain contains only two users and one position, EI struggles to learn diverse domain characteristics.
(iii) RoMF \cite{zhang2024few}, a meta-learning approach, uses a graph neural network (GNN) to measure sample similarity and decouple sensing conditions. However, it achieves only 31.57\% accuracy on our dataset. The primary challenge is that our task involves adapting the HAR model across different users and positions (varying distances and angles), significantly affecting received signals and making it difficult to find cross-condition activity similarities. 
(iv) EADA \cite{xie2022active} and MADA \cite{zhang2024revisiting} active learning methods originally developed for the computer vision domain. EADA selects samples using free energy, while MADA uses Shannon entropy. Overall, active learning methods outperform the other approaches, as they prioritize informative samples for adaptation. Compared to these methods, \system's R$\acute{e}$nyi Entropy-based uncertainty estimation further enhances sample selection, leading to superior performance.

% \begin{table}[h]
% \centering
%   \caption{Ablation study. RE: R$\acute{e}$nyi Entropy. PL: Pseudo Labeling. CL: Contrast Learning.}
%   % \setlength{\tabcolsep}{1.35pt}
%  % \vspace{-0.1in}
%   \begin{tabular}{ccccl}
%     \toprule
%      & Accuracy   &  Precision & Recall  & F1-score \\
%     \hline
%     \textbf{\system~(Ours)}    & 98.03 & 98.10  & 98.03  & 98.03\\
%     \hline
%     w/o RE   &  95.14  &  95.30 &  95.14  &  95.15  \\
%     w/o PL   &  81.48 &86.67  &81.48  &82.91\\
%     w/o CL   &  91.90 &92.01  &91.90  &91.78\\
%   \bottomrule
% \end{tabular}
% % \vspace{-0.15in}
% \label{tab:exp_ablation}
% \end{table}

\subsection{Ablation Study}
To validate the effectiveness of \system's key components, we conduct an ablation study. The results are listed in \Table{exp_overall_comparison}.
`w/o RE': Replacing R$\acute{e}$nyi Entropy with Shannon Entropy for uncertainty estimation leads to degraded performance. This confirms that R$\acute{e}$nyi Entropy effectively focuses on challenging samples, improving adaptation in new domains.
`w/o PL' and `w/o CL': Removing Pseudo Labeling (PL) or Contrastive Learning (CL) leads to a noticeable performance drop. While R$\acute{e}$nyi Entropy identifies the most informative samples for annotation, PL and CL play crucial roles in leveraging unlabeled target data. PL refines model predictions by assigning pseudo labels to unlabeled samples, enabling the model to learn from a larger dataset. CL enhances feature alignment by ensuring that samples of the same activity cluster together in the latent space while pushing apart different activities. 
Moreover, `w/o PLS': Replacing the proposed Pseudo Label Set (PLS) with a single pseudo label results in a more severe performance degradation than removing PL entirely. This suggests that traditional pseudo-labeling strategies are highly susceptible to error accumulation, while our PLS approach significantly mitigates this issue by enhancing robustness during training. Together, these techniques strengthen domain adaptation performance.

\begin{figure}[t]
\centering
\begin{minipage}[h]{0.45\linewidth}
\centering
{\includegraphics[width=\textwidth]{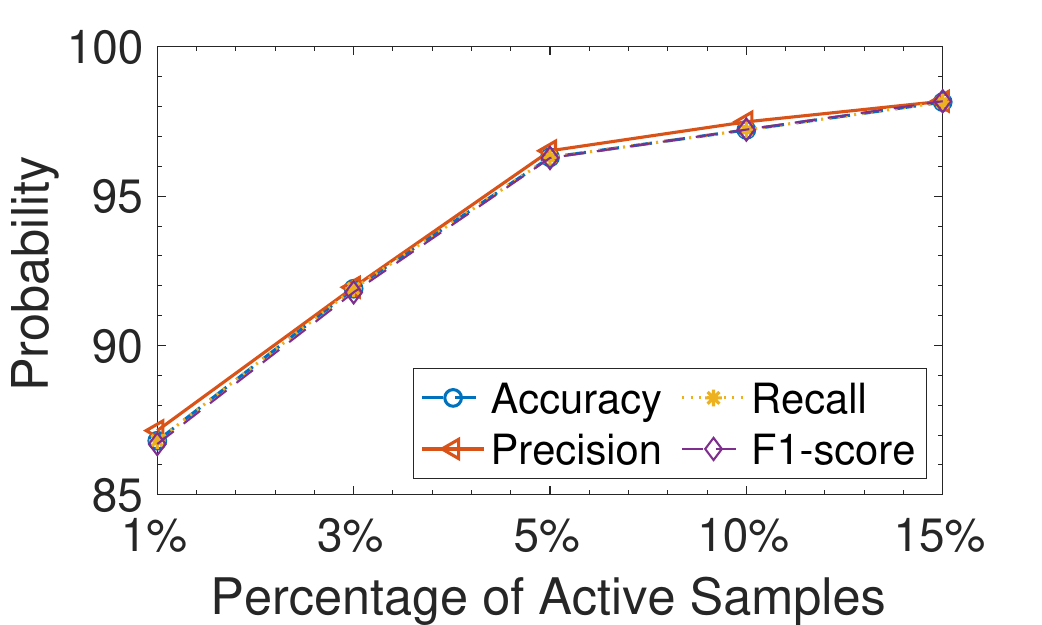}}
\caption{The impact of percentage of active samples.}
\label{fig:exp_budgets}
\end{minipage}
\begin{minipage}[h]{0.45\linewidth}
\centering
\includegraphics[width=\textwidth]{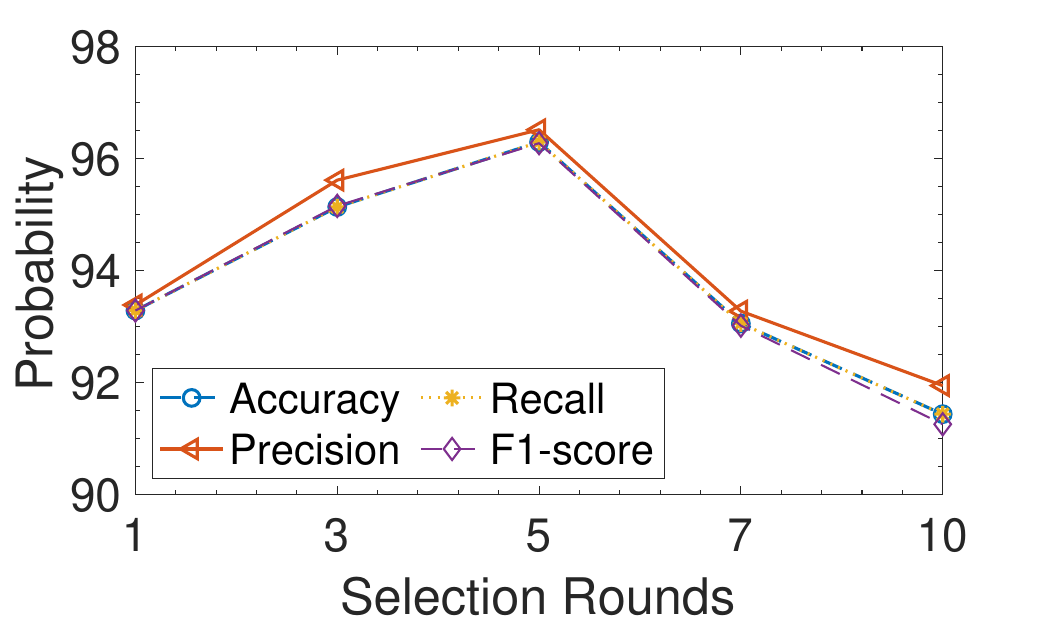}
\caption{The impact of selection rounds.}
\label{fig:exp_selectionRound}
\end{minipage}
\end{figure}

\begin{figure*}[t]
\begin{minipage}[h]{\linewidth}
\centering
\subfigure[Performance vs individuals]
{\includegraphics[width=.25\textwidth]{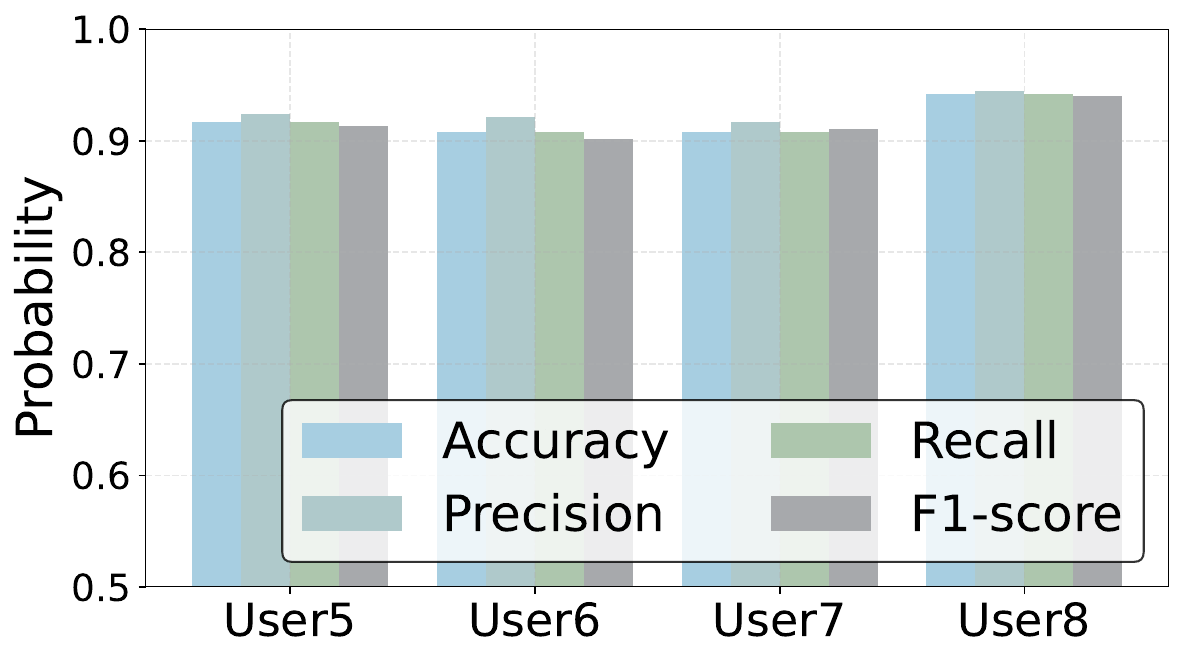}}
\hspace{0.1in}
\subfigure[Performance vs positions]
{\includegraphics[width=.25\textwidth]{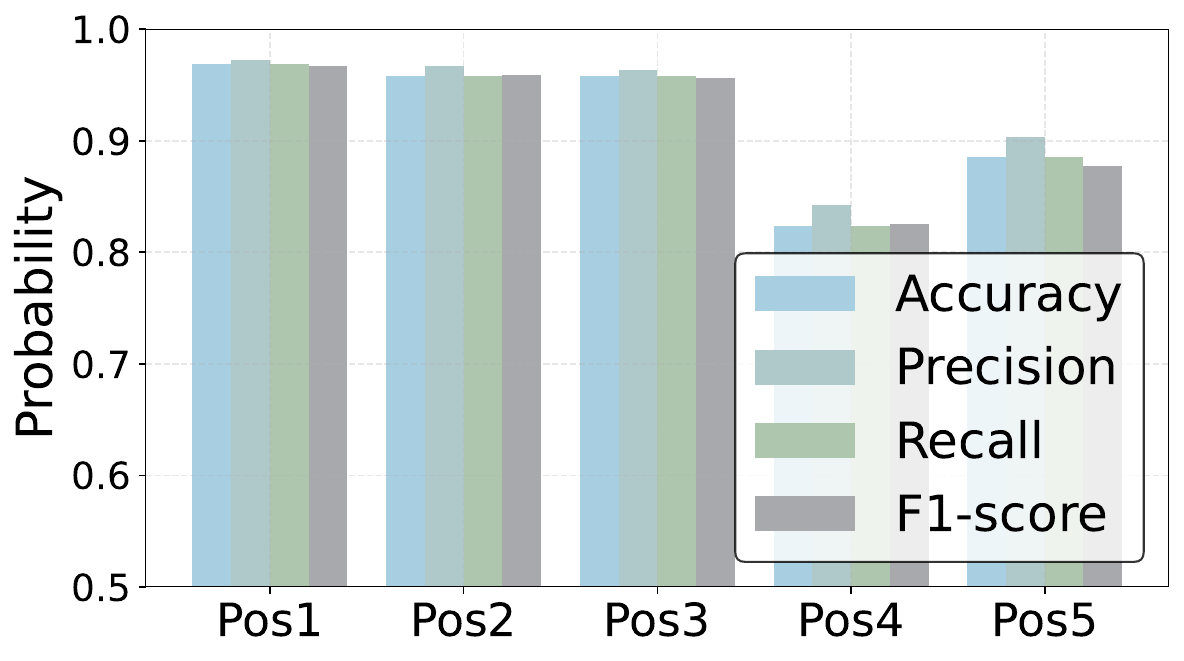}}
\hspace{0.1in}
\subfigure[Shannon vs R$\acute{e}$nyi Entropy]
{\includegraphics[width=.25\textwidth]{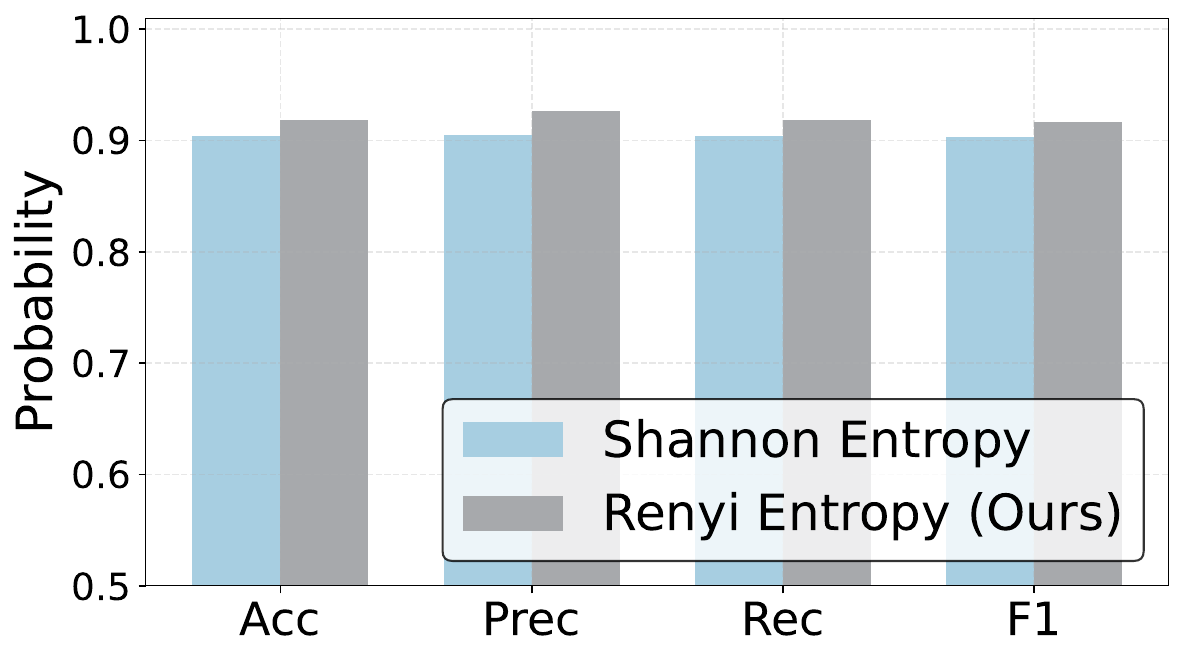}}
\vspace{-0.1in}
\caption{Performance across unseen users.}
\label{fig:exp_newUser}
\end{minipage}
\vspace{-0.1in}
\end{figure*}

\subsection{Micro-Benchmark Experiments}

\subsubsection{Percentage of Active Samples}
By default, we allocate 5\% of the labeling budget, aligning with standard ADA settings. To evaluate the impact of labeling budgets, we vary the percentage from 1\% to 15\% and test \system's performance. Notably, as shown in \Figure{exp_budgets}, \system outperforms baseline methods even with just 1\% of labeled samples (refer to \Table{exp_overall_comparison}), achieving 86.81\% accuracy. As the labeling budget increases, performance gradually improves; however, beyond 5\%, the performance gain slows down.

% \begin{figure*}[h]
% \begin{minipage}[h]{0.49\linewidth}
% \centering
% {\includegraphics[width=0.65\textwidth]{pic/exp_budgets}}
% \caption{The impact of percentage of active samples.}
% \label{fig:exp_budgets}
% \end{minipage}
% % \hspace{0.15in}
% \begin{minipage}[h]{0.49\linewidth}
% \centering
% \includegraphics[width=.65\textwidth]{pic/exp_selectionRound}
% \caption{The impact of selection rounds.}
% \label{fig:exp_selectionRound}
% \end{minipage}
% % \vspace{-0.1in}
% \end{figure*}

\subsubsection{Selection Rounds}
We also examine the effect of selection rounds while keeping the total labeling budget fixed. The results (\Figure{exp_selectionRound}) show that \system achieves best performance with five selection rounds, after which performance starts to decline. This drop is likely due to the limited number of target samples per round, which can cause the model to converge prematurely to a local optimum, hindering overall adaptation. Based on this observation, we set the number of selection rounds to five in our implementation.

\subsection{Generalization Study}
This section evaluates the generalization capability of \system across unseen users, unseen environments, and other open-source datasets.

\subsubsection{Unseen Users}
In this experiment, users from Group \#3 and \#4 serve as unseen participants, meaning their data is not seen during training. They perform the same activities as the seen users in the default scene. We apply the adapted HAR model (the `S1-T1\&2-U3\&4' setting) to recognize their activities directly. As shown in \Figure{exp_newUser}(a), the accuracy ranges from 90.83\% to 94.17\%, with an average of 91.88\%. These results indicate that while activity patterns remain similar, variations in users' physical characteristics and movement styles significantly affect reflected mmWave signals.
Additionally, we analyze the average performance of these unseen users across different positions. As shown in \Figure{exp_newUser}(b), the results are consistent with those of seen users. Pos \#1$\sim$\#3 maintain high accuracy ($>$96\%), while accuracy for Pos \#4 and \#5 drops to approximately 82\% and 89\%.

% \begin{figure*}[t]
% \centering

% \begin{minipage}[t]{\textwidth}
% \centering
% \subfigure[Performance vs individuals]
% {\includegraphics[width=.3\textwidth]{pic/exp_newUser}}
% % \hspace{0.1in}
% \subfigure[Performance vs positions]
% {\includegraphics[width=.3\textwidth]{pic/exp_newUser_diffDomains}}
% \subfigure[Shannon Entropy vs R$\acute{e}$nyi Entropy]
% {\includegraphics[width=.3\textwidth]{pic/exp_newUser_comp}}
% \caption{Performance across unseen users.}
% \label{fig:exp_newUser}
% \end{minipage}
% % \vspace{-0.15in}
% \end{figure*}

\subsubsection{Unseen Environment}\label{sec:proof_renyi_over_shannon}
\begin{figure}[t]
\centering
\begin{minipage}[t]{\linewidth}
\centering 
\subfigure[Unseen environment]
{\includegraphics[width=.49\textwidth]{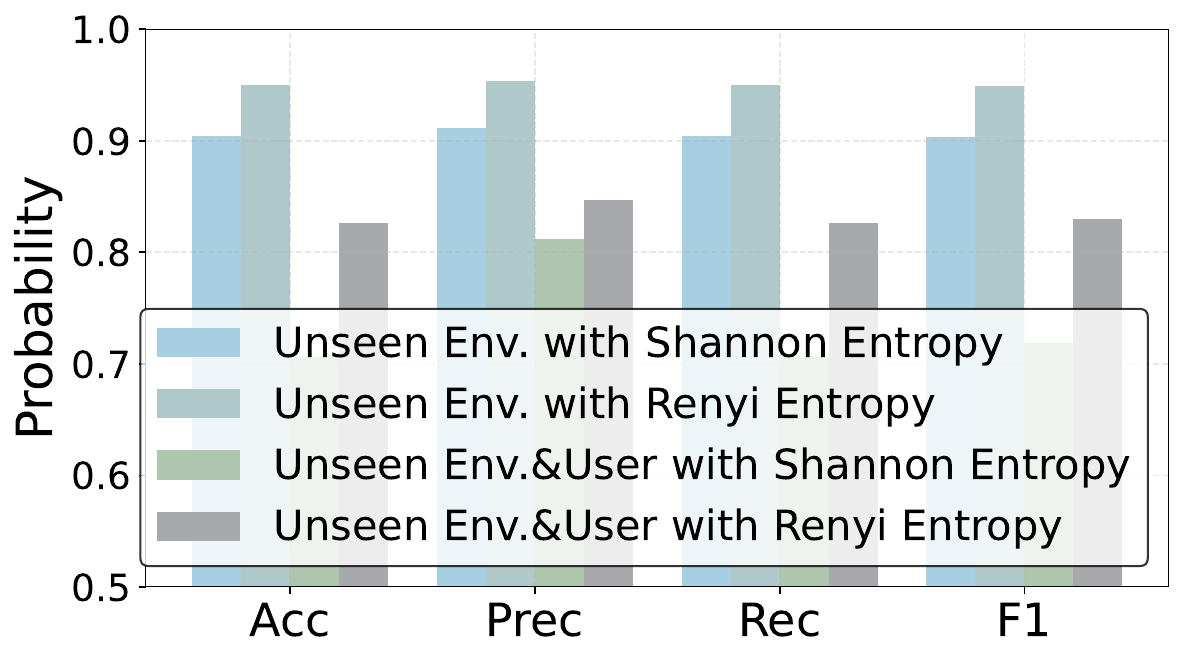}}
% \subfigure[Unseen environment]
% {\includegraphics[width=.245\textwidth]{pic/exp_newEnv}}
% % \hspace{0.1in}
% \subfigure[Unseen environment \& user]
% {\includegraphics[width=.245\textwidth]{pic/exp_newEnvUser}}
\subfigure[Unseen environment \& user]
{\includegraphics[width=.49\textwidth]{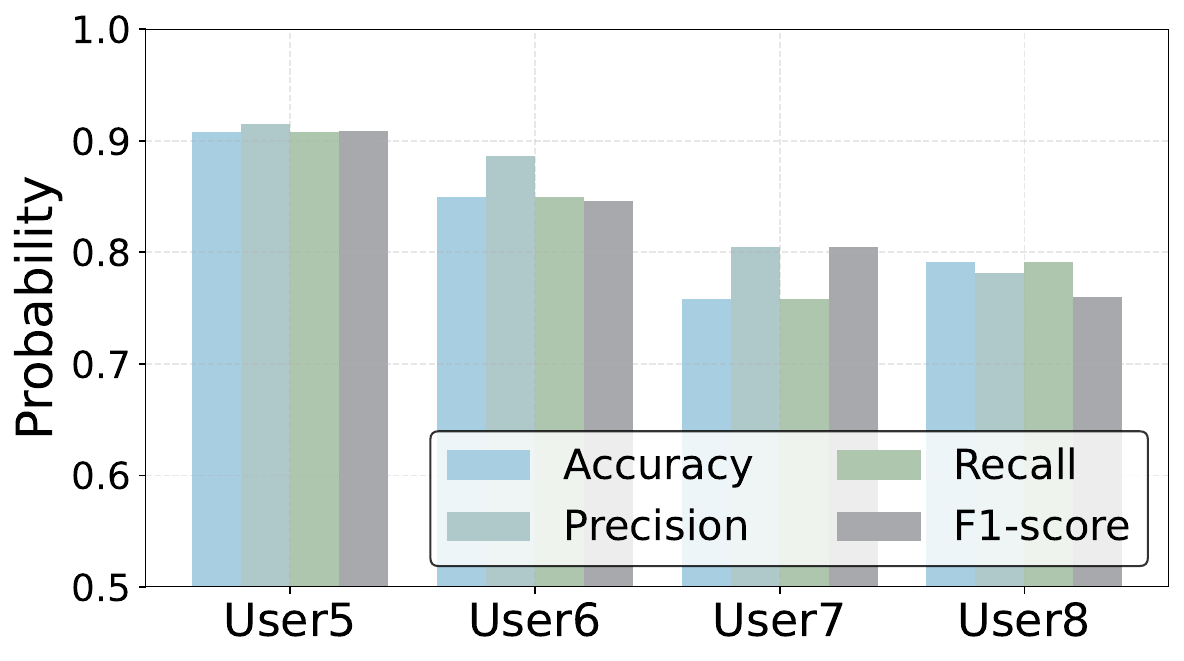}}
\vspace{-0.1in}
\caption{Performance across unseen Env. \& users.}
\label{fig:exp_newEnv}
\end{minipage}
\vspace{-0.1in}
\end{figure}

To assess performance in an unseen environment, we use the model trained in Scene \#1 (\Figure{impl_scene}(a)) to recognize activities in Scene \#2 (\Figure{impl_scene}(b)), which is more complex due to the presence of tables, chairs, and walls nearby the users. When tested on seen users (User \#1–\#4) in Scene \#2, as shown in \Figure{exp_newEnv}(a), the model achieves over 95\% accuracy, suggesting that adapting to a new environment is easier than adapting to new users.
In addition, \Figure{exp_newEnv}(b) presents a more challenging case: unseen users in an unseen environment. Here, the average accuracy drops to 82.71\%, ranging from 90.83\% to 75.83\%, highlighting the increased difficulty of simultaneous user and environment adaptation.

Furthermore, we compare the effectiveness of Shannon Entropy and R$\acute{e}$nyi Entropy. As illustrated in \Figure{exp_newUser}(c) and \Figure{exp_newEnv}(a), R$\acute{e}$nyi Entropy consistently outperforms Shannon Entropy in most cases. 
These results align with \Figure{pre_distribution}, demonstrating that using R$\acute{e}$nyi Entropy for domain and prediction uncertainty estimation leads to more consistent uncertainty distributions across domains. Consequently, the selected labeled samples contribute to improving the model's generalization ability, confirming the advantage of using R$\acute{e}$nyi Entropy in our framework.

\subsubsection{New Datasets}
We further conduct extensive experiments on two additional open-source datasets: XRF55 \cite{wang2024xrf55} and MM-Fi \cite{yang2023mm}.

i) \textbf{XRF55.} XRF55 \cite{wang2024xrf55} is a large-scale multi-modal dataset  focusing on radio frequency (RF) signals. It comprises 42.9K samples collected from 39 human subjects, covering 55 categories of indoor human-object interactions. The dataset provides Range-Doppler (RD) and Range-Angle (RA) heatmaps derived from mmWave data, captured across four distinct scenes (Scene \#1 to \#4).

$\bullet$ \textit{Training and Testing Data.} Each human subject performed every action 20 times within the sensing area. The first 14 trials were designated as training samples, while the remaining 6 trials were used for testing. We conducted two types of experiments to evaluate \system's robustness:
(1) Cross-Person Experiment. Source Domain: data from 20 human subjects (numbered 1-30, excluding multiples of 3) in Scene \#1. Target Domain: data from 10 human subjects (numbered as multiples of 3) in Scene \#1. 
(2) Cross-Scene\&Person Experiment. Source Domain: data from the first 11 human subjects in Scene \#1. Target Domain: data from all 3 human subjects in Scenes \#2, \#3, and \#4. Notably, among these scenes, 2, 2, and 1 subjects, respectively, were unseen users.

$\bullet$ \textit{Implementation Details.} For both experiments, we transform RD and RA heatmaps into Time-Doppler (TD) and Time-Angle (TA) heatmaps, following the method outlined in \Section{process}. We employ ImageNet pretrained ResNet50 \cite{he2016deep} as the feature extractors for both TD and TA branches. The hyperparameters are configured as follows: batch size = 32, $\lambda_{\text {dom}}$ = 7, $\lambda_{\text {pred}}$ = 0.5, $\lambda_c = 0.1$, $k$ = 5, and select $N^B = 5\% N^{Tu}$ target domain samples for every method. 
Additionally, we incorporate the one-shot learning (using approximately 7.14\% of the target domain samples) results of XRF55 into the baseline methods for comparison.

$\bullet$ \textit{Results.} The evaluation results, as shown in \Table{exp_XRF55}, highlight \system's generalization capabilities across individuals and scenes, even on large-scale and challenging datasets. Specifically, in the cross-person experiment, \system improves accuracy from 64.42\% to 70.97\%.
In the cross-scene\&person experiments, \system achieves an average accuracy gain of 33.20\%.
Compared to the one-shot baseline, \system demonstrates a 3.49\% improvement in the cross-person experiment and an average 22.95\% improvement in cross-scene\&person experiments, while using 2.14\% less labeled data in the target domain.
These results underscore \system's ability to generalize effectively across diverse datasets and scenarios, outperforming baselines even with limited labeled data.

% \begin{small}
\begin{table}[t]
  \centering
    \caption{Domain adaptation results on XRF55 dataset.}
  \setlength{\tabcolsep}{1.35pt}
   % \vspace{-0.1in}
    \begin{tabular}{ccccc}
      \toprule
      &&\multicolumn{3}{c}{Cross-scene\&person}\\
      \cmidrule {3-5}
      Accuracy (\%) & Cross-person &  Scene1$\rightarrow$2 & Scene1$\rightarrow$3  & Scene1$\rightarrow$4  \\
      \hline
      \textbf{\system}    & \bf{70.97} & \bf{53.84} & \bf{55.45} & \bf{54.65} \\
      DT                                    & 64.42 &  11.36 & 26.42 & 26.58 \\
      XRF55~\cite{wang2024xrf55} & 67.48 & 25.33 & 34.51 & 35.25 \\
      mTransSee~\cite{liu2022mtranssee}     &  63.65 &  33.20 & 45.70 & 41.91 \\
      EI~\cite{jiang2018towards}  &  18.67 &  8.15 &  15.67 & 1.90 \\
      % RoMF~\cite{zhang2024few}  &  &  &  &    \\
      EADA~\cite{xie2022active}    & 64.79  & \underline{44.34} & \underline{47.88} & \underline{45.56} \\
      MADA~\cite{zhang2024revisiting}   & \underline{67.94} & 36.16 & 45.35 &  43.43 \\
    \bottomrule
  \end{tabular}
  \label{tab:exp_XRF55}
  % \vspace{-0.15in}
  \end{table}
% \end{small}

ii) \textbf{MM-Fi.} MM-Fi \cite{yang2023mm} is the first multi-modal non-intrusive 4D human dataset, consisting of over 320K frames capturing 27 classes of daily activities and rehabilitation exercises. It provides preprocessed point clouds for mmWave data across four distinct environments.

$\bullet$ \textit{Training and Testing Data.} We conduct comprehensive 1 on 1 domain adaptation experiments across the four environments. As mentioned in \cite{yang2023mm}, the number of points in a single original frame (approximately 20) is insufficient for effective activity recognition. To address this, we merge adjacent frames to increase the total number of points to 128. 
Different from \cite{yang2023mm}\cite{an2022fast}, we augment each point in the merged frame with a sixth-dimensional feature—a uniformly distributed position number on the interval $[0, 1]$—indicating its origin frame index. Additionally, the Doppler velocity of each point was normalized by dividing by 100 to ensure numerical stability. For example, in a merged frame combining three original frames, points are represented as: $P_1 = (x_1, y_1, z_1, \frac{D_1}{100}, I_1, 0)$, $P_2 = (x_2, y_2, z_2, \frac{D_2}{100}, I_2, \frac{1}{2})$, and $P_3 = (x_3, y_3, z_3, \frac{D_3}{100}, I_3, 1)$. Here, $(x, y, z)$, $D$ and $I$ denote spatial coordinates, Doppler velocity, and intensity, respectively. The last dimension, \ie, the position number, preserves the temporal origin of each point, ensuring consistency when merging frames of varying lengths. The merged frames are split into training, testing, and validation sets at a 3:1:1 ratio.

% \begin{small}
\begin{table}[t]
\centering
  \caption{Comprehensive 1 on 1 domain adaptation experiment results on MM-Fi dataset.}
  \setlength{\tabcolsep}{1.35pt}
 % \vspace{-0.1in}
  \begin{tabular}{cccccccl}
    \toprule
    Accuracy (\%) & 1$\rightarrow$2   &  1$\rightarrow$3 & 1$\rightarrow$4  & 2$\rightarrow$3  &  2$\rightarrow$4 & 3$\rightarrow$4 & Mean \\
    \hline
    \textbf{\system}    & \bf{97.65} & \bf{97.50} & \bf{93.57} & \bf{99.17} & \bf{96.04} & \bf{97.34} & \bf{96.88}\\
    DT                                    & 64.11 &  60.68 & 31.82 & 73.64 & 53.13 & 54.08 & 56.24\\
    mTransSee~\cite{liu2022mtranssee}     &  81.32 &  77.76 & 78.39 & 81.79 & 73.48 & 70.58&  77.22\\
    EI~\cite{jiang2018towards}  &  65.99 &  65.32 &  61.92 & 78.55&  47.37 & 70.04 & 64.87\\
    RoMF~\cite{zhang2024few}  & 75.55 & 68.04 & 45.30 &  70.25 & 47.02 & 67.08 & 62.21\\
    EADA~\cite{xie2022active}    & \underline{92.01}  & \underline{92.64} & \underline{84.64} & 93.08 & \underline{87.77}&  91.38 & \underline{89.90}\\
    MADA~\cite{zhang2024revisiting}   & 89.97 & 89.99 & 71.00 &   \underline{97.05}&  83.39 & \underline{95.77} & 87.86\\
  \bottomrule
\end{tabular}
\label{tab:exp_mmFi}
% \vspace{-0.1in}
\end{table}
% \end{small}

$\bullet$ \textit{Implementation Details.} We implement the Point Transformer \cite{zhao2021point} pretrained on the source domain as the backbone for feature extraction from point clouds in MM-Fi dataset. 
To ensure uniformity, each merged frame is processed using Farthest Point Sampling (FPS) to sample 128 points, with zero-padding applied if the frame contained fewer points.
During training, the starting point for FPS was randomly selected, while for testing and validation, it was always the first point. Batch size is 16 and other experimental settings mirrored those used for the XRF55 dataset.

$\bullet$ \textit{Results.} As shown in \Table{exp_mmFi}, \system consistently outperforms all baseline methods, achieving a mean accuracy of 96.88\%, significantly higher than the second-best method, EADA (89.90\%). Compared to Direct Transfer (DT), \system achieves an average improvement of 40.64\%.
We also observe that adaptation tasks involving Domain 4 (\eg, 1$\rightarrow$4, 2$\rightarrow$4, 3$\rightarrow$4) are the most challenging, with notable performance drops for most methods, suggesting Domain 4 has distinct characteristics that complicate adaptation. Baselines like DT, EI, and RoMF struggle, with mean accuracies below 65\%, while EADA and MADA show moderate performance. 
These results on MM-Fi dataset highlight \system's effectiveness in adapting to cross-environment scenarios involving daily or rehabilitation activities. 
Furthermore, a comparison of \Table{exp_XRF55} and \Table{exp_mmFi} also confirms that cross-person adaptation is generally more challenging than cross-environment adaptation.

\section{Related Works}

\subsection{Wireless HAR}

HAR has been widely studied, with various techniques developed to detect and classify human actions. Traditional vision-based HAR methods \cite{he2024ma}\cite{shu2024learning}, while achieving high accuracy, are constrained by factors such as lighting conditions, occlusion, and privacy concerns. These limitations have led researchers to explore alternative sensing modalities, giving rise to contactless HAR solutions. Wireless signals, including acoustic signals \cite{cao2023can}\cite{wang2024amt}\cite{ding2022ultraspeech}, Wi-Fi \cite{survey_on_wifi_sensing}\cite{zhao2024one}\cite{feng2022wi}, and RFID \cite{ding2015femo}\cite{zhao2023rfid}\cite{han2015cbid}, have been employed to recognize human activities, offering greater flexibility and reducing privacy concerns. In recent years, mmWave radar has emerged as a promising solution for fine-grained motion profiling, such as human mesh construction \cite{xue2023towards}\cite{ding2023mi}\cite{huang2025one}, vital sign monitoring \cite{wang2022loear}, minor vibration estimation \cite{zeng2023msilent}\cite{dachao2025mmspeech}, and HAR \cite{liu2022mtranssee}\cite{zhao2025federated}\cite{zhao2025mm}. While these mmWave-based systems hold considerable potential, they still face challenges like domain shift, where models trained in one domain fail to perform well when deployed in new domains (users, positions, or sensing conditions).

\subsection{Wireless HAR with Domain Adaptation}
To improve the generalization of wireless HAR systems across various domains, researchers have explored a variety of approaches \cite{survey_on_wifi_sensing}. Some methods, such as Widar3.0 \cite{widar3.0} and EI \cite{jiang2018towards}, aim to extract environment-irrelevant features through signal processing or adversarial learning techniques. Other approaches use fine-tuning-based transfer learning for domain adaptation. For instance, mTransSee \cite{liu2022mtranssee} utilizes few-shot learning to adapt the activity classifier to new environments. 
Similarly, RF-Net \cite{ding2020rf} employs a meta-learning framework to fine-tune a pre-trained model for unseen conditions.
Additionally, RoMF \cite{zhang2024few} constructs a graph to profile the similarity between activities, using graph neural networks (GNNs) and meta-learning to adapt the HAR system to various conditions. 
Some methods also leverage multi-modal data, such as video \cite{xia2024ts2act}, text \cite{weng2024large}, and IMUs \cite{bhalla2021imu2doppler}, to enhance domain adaptation capabilities. 
In contrast to these existing methods, we propose a novel solution that employs active learning to address the domain shift problem in wireless HAR. Our method focuses on selecting and labeling a limited number of highly informative target domain samples, minimizing retraining efforts while enhancing system robustness in dynamic conditions. 
% This approach not only reduces annotation costs but also enhances the generalization of HAR systems across different conditions.

\subsection{Active Domain Adaptation}

Active Domain Adaptation (ADA) enables the adaptation of a source model to a new domain by labeling a small number of target samples. AADA \cite{su2020active} applies a domain discriminator to evaluate the domain characteristics of target samples and weighs them based on entropy-based uncertainty.
CLUE \cite{fu2021transferable} performs entropy-weighted clustering to select uncertain and diverse samples. 
%
% SDM [15] uses a unique margin loss for training and selects data near the margin. 
%
EADA \cite{xie2022active} employs a free energy bias to measure the differences between the source and target domains, minimizing domain discrepancies by aligning energies.
Although these methods have achieved empirical success, they primarily focus on compensating the domain gaps by incorporating target domain representations into the query function. However, uncertainty measures they use are still based on point estimates from deterministic models, which can be miscalibrated when dealing with out-of-distribution data.
To address this, DiaNA \cite{huang2023divide} uses a Gaussian Mixture Model to measure the uncertainty of candidate samples and selects them accordingly. 
DUC \cite{xie2023dirichlet} uses Dirichlet priors and Shannon entropy on class probabilities to evaluate domain and prediction uncertainty.
Building on DUC, MADA \cite{zhang2024revisiting} improves sample selection by balancing uncertainty with sample diversity, achieving better results. 
Shannon entropy can be seen as R$\acute{e}$nyi entropy with order $s = 1$. When $s < 1$, the entropy of the same distribution increases, which allows R$\acute{e}$nyi entropy to amplify the differences in uncertainty between samples. This property helps the model focus on samples with higher uncertainty during training and sample selection, leading to better performance. Hence, in this paper, we explore R$\acute{e}$nyi entropy for uncertainty estimation, and demonstrate that it yields better results across a variety of conditions and datasets.

\section{Conclusion}

This paper introduces \system, an active domain adaptation framework designed to improve the generalization of mmWave-based HAR models across agnostic new domains, such as different users, positions, and environments, while minimizing labeled data requirements. To achieve this, \system incorporates R$\acute{e}$nyi Entropy-based uncertainty estimation to select the most informative target samples for adapting, outperforming traditional metrics like Shannon Entropy. Comparisons with five representative baselines confirm its effectiveness. Further experiments on unseen users, environments, and two open-source datasets validate its robustness and generalization capability.
Comprehensive results also indicate that cross-person and cross-angle adaptation remain particularly challenging for mmWave-based HAR. This highlights the need for further research into advanced signal processing and deep learning techniques to enhance system robustness in diverse real-world scenarios.

%\subsection*{Acknowledgements}
%We would like to thank ...

% \newpage
% \small{
\bibliographystyle{IEEEtran}
\bibliography{refs}
% }

% \newpage
% \appendix
% \setcounter{equation}{0}
% \renewcommand{\theequation}{A\arabic{equation}}
% \input{Body/appendix.tex}

\end{document}